\documentclass{jfm}
\usepackage{graphicx}
\usepackage{mathtools}
\usepackage{epstopdf, epsfig}
\usepackage{tikz}
\usepackage[ruled]{algorithm2e}
\usepackage{booktabs}
\usepackage{array}
\usepackage{amsmath, amssymb}
\usepackage{multirow}
\usepackage{comment}
\usepackage{natbib}
\usepackage[]{color}
\usepackage[]{xcolor}
\usepackage{placeins}
\usepackage{relsize}
\usepackage{soul}
\usepackage{colortbl}
\usepackage{upgreek}
\usepackage{cleveref}
\usepackage{url}
\usepackage[normalem]{ulem}

\newcommand{\solidLine}[3]{%
  \tikz[baseline=-0.5ex]{%
    \draw[draw={rgb,1:red,#1;green,#2;blue,#3}, thick] (-0.9ex,0) -- (0.9ex,0);
  }%
}

\newcommand{\dashedLine}[3]{%
  \tikz[baseline=-0.5ex]{%
    \draw[draw={rgb,1:red,#1;green,#2;blue,#3}, thick, dashed] (-0.9ex,0) -- (1.3ex,0);
  }%
}

\newcommand{\dottedLine}[3]{%
  \tikz[baseline=-0.5ex]{%
    \draw[draw={rgb,1:red,#1;green,#2;blue,#3}, thick, dotted, line width=1.5pt] (-0.7ex,0) -- (0.9ex,0);
  }%
}

\newcommand{\dashedDottedLine}[3]{%
  \tikz[baseline=-0.5ex]{%
    \draw[draw={rgb,1:red,#1;green,#2;blue,#3}, thick,
      dash dot, line width=1.0pt] (-0.9ex,0) -- (1.9ex,0);
  }%
}

\newcommand{\colorcircle}[6]{%
  \tikz[baseline=-0.5ex]{%
    \filldraw[
      fill={rgb,1:red,#1;green,#2;blue,#3},
      draw={rgb,1:red,#4;green,#5;blue,#6},
      line width=1pt
    ] (0,0) circle (0.5ex);
  }%
}

\newcommand{\colortriangle}[6]{%
  \tikz[baseline=-0.5ex]{%
    \filldraw[
      fill={rgb,1:red,#1;green,#2;blue,#3},
      draw={rgb,1:red,#4;green,#5;blue,#6},
      line width=1pt
    ] 
    (0,0.5ex) -- (-0.5ex,-0.5ex) -- (0.5ex,-0.5ex) -- cycle;
  }%
}

\newcommand{\colortriangleright}[6]{%
  \tikz[baseline=-0.5ex]{%
    \filldraw[
      fill={rgb,1:red,#1;green,#2;blue,#3},
      draw={rgb,1:red,#4;green,#5;blue,#6},
      line width=1pt
    ] 
    (-0.5ex,0.0ex) -- (0.5ex,-0.5ex) -- (0.5ex,0.5ex) -- cycle;
  }%
}

\newcommand{\colortriangleleft}[6]{%
  \tikz[baseline=-0.5ex]{%
    \filldraw[
      fill={rgb,1:red,#1;green,#2;blue,#3},
      draw={rgb,1:red,#4;green,#5;blue,#6},
      line width=1pt
    ] 
    (0.5ex,0.0ex) -- (-0.5ex,-0.5ex) -- (-0.5ex,0.5ex) -- cycle;
  }%
}

\newcommand{\colorplus}[3]{%
  \tikz[baseline=-0.5ex]{%
    \draw[
      draw={rgb,1:red,#1;green,#2;blue,#3},
      line width=1.0pt
    ] (-0.8ex,0) -- (0.8ex,0)
      (0,-0.8ex) -- (0,0.8ex);
  }%
}

\newcommand{\colorcross}[3]{%
  \tikz[baseline=-0.5ex]{%
    \draw[
      draw={rgb,1:red,#1;green,#2;blue,#3},
      line width=1.0pt,
      line cap=round
    ] (-0.6ex,-0.6ex) -- (0.6ex,0.6ex)
      (-0.6ex,0.6ex) -- (0.6ex,-0.6ex);
  }%
}

\newcommand{\colorsquare}[6]{%
  \tikz[baseline=-0.5ex]{%
    \filldraw[
      fill={rgb,1:red,#1;green,#2;blue,#3},
      draw={rgb,1:red,#4;green,#5;blue,#6},
      line width=1pt
    ]
    (-0.5ex,-0.5ex) rectangle (0.5ex,0.5ex);
  }%
}

\usetikzlibrary{shapes.geometric}
\newcommand{\colorstar}[6]{%
  \tikz[baseline=-0.5ex]{%
    \node[
      star,
      star points=5,
      star point ratio=2.25,
      minimum size=6pt,
      inner sep=0pt,
      fill={rgb,1:red,#1;green,#2;blue,#3},
      draw={rgb,1:red,#4;green,#5;blue,#6},
      line width=0.8pt
    ] {};
  }%
}

\newcolumntype{x}[1]{>{\centering\arraybackslash\hspace{0pt}}p{#1}}

\def\del#1{\textcolor{gray}{~}} 

\allowdisplaybreaks[1]

\shorttitle{Assimilation of measurements in DNS of high-speed flow over a cone-flare}
\shortauthor{P.\,Morra, B.\,Tillman, S.\,Laurence \& T.\,A.\,Zaki}
\title{Assimilation of wall-pressure measurements in direct numerical simulations of high-speed flow over a cone-flare geometry}

\author{Pierluigi Morra\aff{1}, 
Brett Tillman\aff{1},
Stuart Laurence\aff{2} \\
\and Tamer A.\,Zaki\aff{1,3}   \corresp{\email{t.zaki@jhu.edu}}}

\affiliation{\aff{1}Department of Mechanical Engineering, Johns Hopkins University,
Baltimore, MD 21218, USA
\aff{2}Department of Aerospace Engineering, University of Maryland, College Park, MD 20742, USA
\aff{3}Department of Applied Mathematics \& Statistics, Johns Hopkins University,
Baltimore, MD 21218, USA}
\begin{document}

\maketitle

\begin{abstract}
Ensemble-variational (EnVar) assimilation of wall-pressure measurements in direct numerical simulations of Mach 6 flow over a cone-flare is performed. The experimental data include pressure spectra and intensities from seven wall-mounted PCB sensors positioned upstream, within, and downstream of the separation region induced by the compression corner. 
Assimilation of the first two sensors only, all upstream of separation, is insufficient to accurately predict the downstream flow.  
Assimilating all the sensor data is shown to be essential to correctly predict separation onset and the downstream wall-pressure data. 
Similar to the experiments, the assimilated flow features intense rope-like structures in the attached region.  
The simulations additionally predict a localized amplification of disturbances beneath the separation shock, where experimental data are not available.
This amplification results from the interaction of the boundary-layer instability modes with the compression shock.  The simulations also capture the sharp decrease in wall-pressure intensity across separation, and the amplification of low-frequency three-dimensional disturbances within the recirculation bubble.
Additionally, the computations highlight the uncertainty in the post-separation predictions due to the low-frequency unsteadiness of the separation shock.  Oscillations of the streamwise velocity modulate the boundary-layer thickness, which in turn introduces variability in disturbance amplification.  
\end{abstract}


\section{Introduction}
\label{sec:introduction}

\begin{figure}
    \centering
    \includegraphics[width=\textwidth]{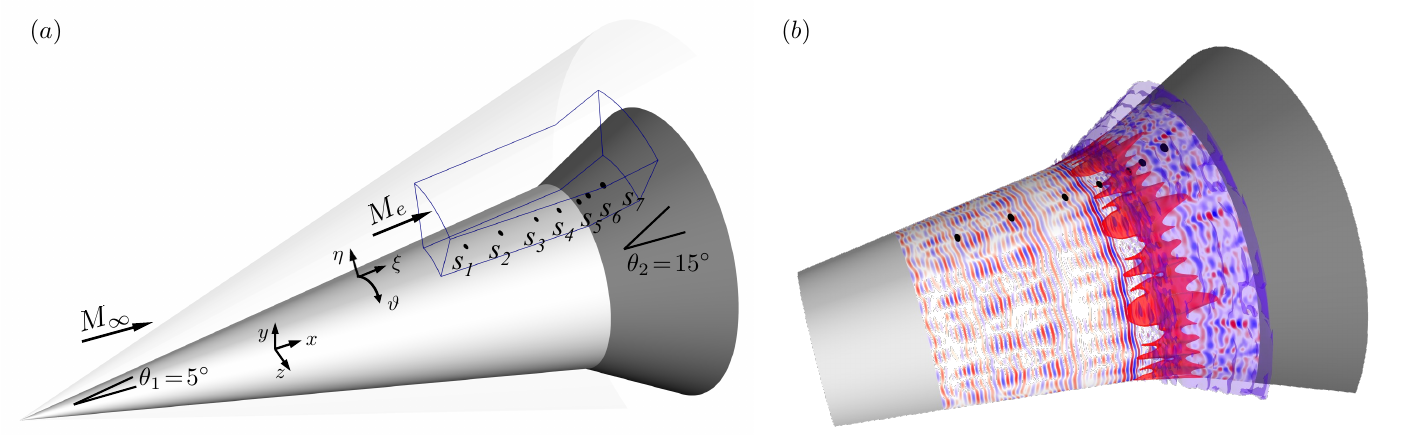}
    \caption{Flow configuration. (\textit{a}) Cone-flare geometry, computational domain, and sensor locations. (\textit{b}) Snapshot of the simulated flow. Blue-white-red contours show wall-pressure fluctuations ($p^{\prime}$); red iso-surface marks separation identified as by zero streamwise velocity ($u_{\xi}=0$); purple iso-surface is the corner shock. (a,b) The axial length is scaled down by a factor of two.}
    \label{fig:fig01}
    \vspace*{-4pt}
\end{figure}

Accurate numerical prediction of the true state of a high-speed transitional boundary layer is challenging due to the flow sensitivity to uncertain environmental conditions. Simulations often attempt to address this uncertainty by relying on trial-and-error approaches to model environmental factors through boundary conditions. However, this method is inefficient and can lead to simulations that fail to accurately represent the flow of interest. Data assimilation (DA) provides a rigorous alternative by infusing simulations with experimental measurements. In the present study, an ensemble variational (EnVar) DA technique is applied to simulate the high-speed flow over a cone-flare geometry, with the assimilated measurements taken from a limited number of high-frequency pressure sensors mounted on the surface near the compression corner (figure \ref{fig:fig01}). 

High-speed boundary layer flows over cone-flare geometries received attention as early as the 1950s, with \cite{becker1956_transfer} reporting a \textit{`serious heating problem'} attributed to \textit{`the occurrence of shock--boundary-layer interaction with separation'}. This shock--boundary-layer interaction (SBLI) arises as the compression corner generates an adverse pressure gradient, separating the flow and leading to the development of a recirculation bubble together with an oblique shock upstream of separation. The experiments by \cite{chapman1957_investigation}, \cite{schaefer1962_investigation}, \cite{ginoux1965_investigation} and \cite{heffner1993_experimental} also reported an intensification of the thermomechanical surface loading in presence of a compression corner. Additionally, \cite{ginoux1965_investigation} and \cite{heffner1993_experimental} observed three-dimensional patterns in the form of streaky motions across the reattachment region of the recirculation bubble. 

Since the flow on the cone-flare geometry is not parallel and the shock produces nonlinear interactions with the boundary layer, existing theoretical results from simpler configurations provide a partial view only for how disturbances may behave in such situations.  Those simpler configurations include, for example, the classical free-stream disturbance--shock interactions \citep{ribner1953_convection, kovasznay1953_turbulence, mckenzie1968_interaction} or parallel and spatial boundary layers without shocks \citep{mack1975_linear, mack1984_boundary} which are relevant upstream of the compression corner.  
Numerical studies with high spatio-temporal resolution can provide a window into the dynamics of these flow fields \citep{pruett1998_direct, pagella2001_numerical, lawal2001_direct, adams2001_transition, vandomme2003_transitional}, and have reported intensification of thermomechanical loads across and downstream of the reattachment point of the shear layer on the flare. 
\cite{pagella2001_numerical} quantified prediction errors of linear stability analysis (LSA) based on the parallel flow assumption, and \cite{adams2001_transition} documented stabilization of the most unstable mode predicted by linear stability theory of \cite{mack1984_boundary}, the second Mack mode, within the recirculation bubble of a Mach 5 boundary layer over a flat plate with a $+15^{\circ}$ compression ramp. Simulations by \cite{balakumar2005_stability} for Mach 5.373 flow over a $+5.5^{\circ}$ compression corner, emulating the Hyper-X model \citep{berry2000_forced}, also revealed that the second Mack mode amplifies prior to separation, stabilizes within the recirculation bubble, and amplifies again downstream of reattachment. 
Experiments on axisymmetric geometries with compression corners at different Reynolds and Mach numbers have shown that configurations in which the flow transitions near reattachment can exhibit three to five times the heat transfer of fully turbulent cases, suggesting that triggering transition to turbulence upstream of the separation point might mitigate material loading \citep{benay2006_shock, bur2009_experimental, estruchsamper2012_axisymmetric, estruchsamper2013_effect, vanstone2013_shock}. This highlights the importance of analyzing transitional boundary layers.

Upstream of separation, the disturbance dynamics generally align with linear stability theory \citep{mack1984_boundary, malik1991_stability, kennedy2018_investigation}, showing the emergence of the second Mack mode as the most unstable and dominant feature. However, streaky patterns are also observed in numerical simulations of flared or straight cones without compression corners \citep{laible2011_hypersonic, hader2017_resonance, hader2018_natural, kennedy2022_instability, buchta2022_cone}, and can be generated by multiple mechanisms.  The streaks can arise within boundary layers from nonlinear interactions of oblique waves \citep{thumm1991_transition, fasel1991_breakdown, fasel1993_oblique} or due to non-modal amplification as reported by \cite{caillaud2025_separation} for a cone-cylinder-flare configuration at Mach $6$.

\cite{lugrin2021_transition} performed direct simulations and linear analysis of Mach $5$ flow on a cylinder-flare with a $+15^{\circ}$ compression corner.  They underscored the role of oblique modes for the nonlinear amplification of streaks, upstream and within the separation region.
\cite{paredes2022_cone} performed simulations of a separated cone-cylinder-flare geometry, and showed that oblique convective disturbances experience larger amplification rates than planar disturbances on the separated shear layer: these oblique waves also likely correspond to the shear-layer disturbances observed in the experiments of \cite{butler2022_coneflare} and \cite{benitez2020_instability}. The results reveal two scenarios: (1) oblique waves dominate upstream of separation, or (2) oblique waves emerge after separation as the initially dominant second Mack modes (planar) are dampened and stabilized by the recirculation bubble. Since oblique waves generate streaky structures through nonlinear interactions, these results help to explain the ubiquity of streaky structures in such flows.

Numerical simulations and the global linear stability analysis conducted by \cite{sidharth2018_doublewedge} on a flat plate with a double wedge at Mach 5 identified a critical compression angle beyond which global unstable modes in the form of streaks emerge. That study also concluded that the streaky instability does not originate from centrifugal effects such as {G\"ortler} vortices, as had earlier been hypothesized. \cite{esquieu2019_flow} performed linear stability analysis for a cone-cylinder-flare configuration in a Mach 6 boundary layer and compared the results with experiments. Their findings demonstrated that linear computations align with experimental observations to some extent, confirming the presence of linear growth within the recirculation bubble. The input-output linear analysis by \cite{dwivedi2019_reattachment} for a Mach-8 boundary layer flow over a flat plate with a $+15^{\circ}$ compression corner suggests that the amplification of streaks within the recirculation bubble is primarily driven by baroclinic effects. 
\cite{song2025_instabilities} analyzed a Mach 6 boundary layer over a flat plate with different compression corners. They noted that the amplification of streaky structures and oblique waves in the separated region follow similar trends, and suggested a correlation with curvature and a possible role of the {G\"ortler} mechanism at separation and reattachment.

The experiments by \cite{butler2021_secondmode,butler2022_coneflare} on a Mach-6 boundary layer over a cone-flare geometry captured a number of interesting physical phenomena, such as the rapid growth of shear-layer disturbances within the separated region that could undergo spontaneous radiation of energy, and the propagation of second-mode disturbance energy along the separation and reattachment shocks. In the quiet-tunnel experiments of \cite{benitez2023_shearlayer}, similar shear-layer disturbances were found to break down into turbulent spots within the reattached boundary layer. Complementing these latter experimental findings, the linear stability analysis by \cite{paredes2022_cone} on the same cone-cylinder-flare geometry identified global unstable modes in the form of streaky structures as the most unstable features within the recirculation bubble of the axisymmetric geometry, but could not capture the full nonlinear dynamics behind the experimental data used for comparison. These results further reinforce the value of a quantitative approach to interpret limited experimental measurements, which we achieve using data assimilation.

Data assimilation combines experiments and simulations to enable the prediction of flow fields beyond the scope of direct measurements \citep{zaki2025arfm}. Given experimental data, DA involves solving the inverse problem of determining the input parameters for simulations in order to reproduce the measurements. This approach allows the retrieval of the flow state corresponding to the experimental data, providing non-intrusive access to all flow quantities.
Numerous DA strategies have been developed and applied in fluid dynamics, including filtering and nudging 
\citep{clarkdileoni2020_synchronization,wang2022_synchronization}, 
adjoint and ensemble-variational methods \citep{zaki2021_reconstruction,wang2022_observable,mons2019_kriging}, and neural networks 
\citep{buzzicotti2021_reconstruction,clarkdileoni2023_neural,hao2023_instability,morra2024_ml}.
For example, using wall-pressure measurements, \cite{buchta2022_cone} successfully applied an ensemble-variational (EnVar) technique to compute the energy spectral makeup of incoming external disturbances in a Mach-6 boundary layer over a cone. Their analysis enhanced the accuracy of interpreting sensor data within the nonlinear, transitional flow regime.

Using experimental wall-pressure measurements from flow over a cone-flare geometry (specifically, those from \citealt{butler2022_coneflare}), the present work aims to estimate the upstream boundary-layer disturbances that reproduce the sensor data. 
The energy spectra of the oncoming disturbances are identified, enabling a full simulation of the flow field. This approach leverages the relationship between wall-pressure observations and the upstream flow \citep{wang2025DOD}, and yields a reconstructed flow field that surpasses the resolution of the measurements and permits a deeper analysis of the flow phenomena. 
The importance of sensors within separation for accurate estimation of the upstream flow is highlighted, followed by a detailed analysis of the disturbance fields in the  attached boundary layer and across the separation shock. The influence of the unsteady shock-boundary layer interaction on the uncertainty of the estimation is also examined.

The original experimental configuration and the computational setup are introduced in \S\ref{sec:config}. The experimental measurements and the ensemble variational (EnVar) DA algorithm are presented in \S\ref{sec:da}, along with the optimization protocol. 
The DA solutions with partial observations and with all the sensor data are presented in \S\ref{sec:results_coarse} and \S\ref{sec:results_fine}. 
The disturbance dynamics across the shock-boundary-layer interaction, onset of separation, and in the reattached flow are the subjects of \S\ref{sec:results_shock}-\S\ref{sec:results_mismatch}, and conclusions are provided in \S\ref{sec:conclusion}.

\section{Flow configuration}
\label{sec:config}
%
\subsection{Experimental parameters}
\label{sec:config_exp}

The present study focuses on assimilation of available experimental measurements of high-speed flow over a cone-flare in direct numerical simulations.  While the focus is on the assimilation task and the results from the simulations, we include a brief description of the relevant experimental parameters for completeness \citep[for details, see][]{butler2022_coneflare}. 

The experiment was conducted in the HyperTERP reflected-shock tunnel at the University of Maryland \citep{butler2019_hyperterp, butler2021_secondmode}. The facility consists of a reflected shock tube to generate the effective stagnation conditions, a converging-diverging nozzle to accelerate the flow to hypersonic Mach numbers, a test section where the test article is housed, and a dump tank. Table~\ref{tab:constants} reports the free-stream fluid properties upstream of the test article, including the free-stream density $\rho_{\infty}$, pressure $p_{\infty}$, temperature $\Theta_{\infty}$ , speed of sound $a_{\infty}$, velocity $U_{\infty}$, Mach number $M_{\infty}= U_{\infty}/a_{\infty}$, dynamic shear viscosity $\mu_{\infty}=\mu_{\scriptscriptstyle \Theta=100\mathrm{K}}(\Theta_{\infty} / 100\mathrm{K})^{0.8497}$, and unit Reynolds number $Re_{\infty}/L =  \rho_{\infty}U_{\infty}/\mu_{\infty}$, where $L$ is a reference length. 
The exponent $0.8497$ for the shear viscosity as function of temperature results from fitting the power law to experimental data from \cite{touloukian1975_thermophysical}, with $\mu_{\scriptscriptstyle \Theta=100\mathrm{K}} = 7.06\times 10^{-6}\,\mathrm{[kg\cdot s /m]}$ being the viscosity at $100\,\mathrm{K}$.
The table also reports the corresponding stagnation density $\rho_{0}$, pressure $p_{0}$, temperature $\Theta_{0}$,  and speed of sound $a_{0}$.

\begin{table}
  \centering
  \begin{tabular}{lccccccccc}
    &
    $\rho$ & 
    $p$ & 
    $\Theta$ &
    $a$ & 
    $U$ & 
    $M$ & 
    $\mu$&
    $Re_{a}/L$&
    $Pr$\\
    &
    $\mathrm{[kg/m^3]}$ & 
    $\mathrm{[Pa]}$ & 
    $\mathrm{[K]}$ & 
    $\mathrm{[m/s]}$ & 
    $\mathrm{[m/s]}$ & 
     & 
    $\mathrm{[kg\! \cdot \! s/m]}$ &
    $\mathrm{[1/m]}$\\[0.5em]
     Stagnation $(\cdot)_{0}$ &
     $5.28$  & 
     $1.35\! \times \! 10^{6}$&
     $890$&
     $598$&
     --- &
     ---&
     ---&
     ---&
     ---\\
     Free-stream $(\cdot)_{\infty}$&
     $2.74 \! \times \! 10^{-2}$  & 
     $855$&
     $109$&
     $209$&
     $1253$&
     $6.00$&
     $7.57 \! \times \! 10^{-6}$&
     $7.56 \! \times \! 10^{5}$&
     ---\\
     Edge $(\cdot)_{e}$&
     $3.74 \! \times \! 10^{-2}$  & 
     $1321$&
     $123$&
     $222$&
     $1242$&
     $5.59$&
     $8.41 \! \times \! 10^{-6}$&
     $9.88 \! \times \! 10^{5}$&
     0.72\\
  \end{tabular}
  \caption{
  Stagnation and free-stream tunnel conditions (\S\ref{sec:config_exp}), and boundary-layer-edge conditions at $x=29.86\,\mathrm{cm}$ from the cone nose tip (\S\ref{sec:config_comp}). Stagnation, free-stream, and edge conditions are denoted with subscripts $\{0, \infty, e\}$.}
  \label{tab:constants}
\end{table}
\begin{table}
  \centering
  \begin{tabular}{lcccccccc}
   Sensor & $s_1$ & $s_2$ & $s_3$ & $s_4$ & $s_5$ & $s_6$ & $s_7$ & $s_{8}$ \\[0.5em]
   Location, $x_{s_i}$ $\mathrm{[cm]}\quad$ & $~32.48~$ & $~35.17~$ & $~37.96~$ & $~39.75~$ & $~41.33~$ & $~42.03~$ & $~43.17~$ & $~44.42~$\\
  \end{tabular}
  \caption{Sensors locations along the $x$-axis. Sensor $s_8$ data are not assimilated, and will be used for independent validation.}
  \label{tab:sensors}
\end{table}

The test article is a cone-flare geometry with a $10^{\circ}$ compression corner, and is shown schematically in figure~\ref{fig:fig01}.  The cone has a circular cross-section, a sharp nose with $0.01\,\mathrm{cm}$ tip radius, a $5^{\circ}$ half-angle, and a $41.00\,\mathrm{cm}$ length. The flare is a circular frustum with a $15^{\circ}$ half-angle and a $7.62\,\mathrm{cm}$ length. Throughout, lengths are measured along the axis of revolution from the nose unless otherwise specified.  The test article is oriented at a zero angle of incidence with respect to the streamwise direction. The model is at room temperature ($\Theta_{w}=300\,\mathrm{K}$) before the flow is accelerated to hypersonic conditions. The test time ($6\,\mathrm{ms}$) is sufficiently brief that changes in the surface temperature can be neglected.
Nine PCB 132B38 pressure transducers are flush-mounted on the surface of the cone-flare, following the line of the axis of revolution. The data assimilation considered the first seven sensors, which comprise of all four sensors on the cone, designated $s_1$ though $s_4$, and the first three sensors on the flare, $s_5$ through $s_7$.  The downstream positions of these sensors along the cone axis are listed in table~\ref{tab:sensors}. Sensors downstream of $s_7$ are excluded, as we focus on the region upstream of turbulence where the dynamics is transitional and dominated by unsteady disturbances interacting with the boundary layer, shocks, and the recirculation bubble. 
Data from sensor $s_8$ will, however, be used for independent comparison to the results from the assimilation of the data from the preceding seven sensors.
A description of the experimental measurements used for data assimilation is provided in \S\ref{sec:da_observations}.

\subsection{Computational setup}
\label{sec:config_comp}

The data assimilation requires a computational model, which in the present study is direct numerical simulation of the compressible Navier-Stokes equations.  The simulation domain is the sub-volume marked in figure \ref{fig:fig01}, downstream of the conical shock which has an angle $\theta_s = 10.64^{\circ}$.  
The edge conditions are determined using the free-stream experimental values from \S\ref{sec:config_exp}, the thermodynamic relations across the leading shock, and the Taylor-Maccoll approximation for inviscid axisymmetric flow above the boundary layer of a cone \citep{stewartson1964_theory,malik1991_stability}. The edge conditions are summarized in table~\ref{tab:constants}.

Two coordinate systems will be adopted where convenient: Cartesian coordinates $(x, y, z)$ and curvilinear body-fitted coordinates $(\xi, \eta, \vartheta)$, in a reference frame centered at the nose tip of the cone-flare. 
In the Cartesian frame, which is marked on figure \ref{fig:fig01}, the cone-flare axis of revolution lies along the $x$-axis, the $xy$-plane contains both the axis of revolution and the sensors, and the $z$-axis is perpendicular to the $xy$-plane.  
In the curvilinear frame of reference, the $\xi$-axis runs parallel to the cone-flare wall, the $\eta$-axis is normal to the cone wall, and the $\vartheta$-axis is perpendicular to the $\xi\eta$-plane.

The fluid is assumed to be a calorically perfect gas with a ratio of specific heats $\gamma = C_p/C_v = 1.4$. The flow is governed by the dimensional Navier-Stokes equations,
\begin{gather}
    \frac{\partial \rho}{\partial t} + \frac{\partial (\rho u_j)}{\partial x_j} = 0,\label{eq:NS1}\\
    \frac{\partial (\rho u_i)}{\partial t} + \frac{\partial}{\partial x_j}(\rho u_i u_j) = - \frac{\partial p}{\partial x_i} + \frac{\partial \tau_{ij}}{\partial x_j},\label{eq:NS2}\\
    \frac{\partial E}{\partial t} + \frac{\partial}{\partial x_j} [u_j(E+p)] = \frac{\partial }{\partial x_j} (u_i \tau_{ij}) + \frac{\partial }{\partial x_j} \left( \kappa \frac{\partial \Theta}{\partial x_j}\right), \label{eq:NS3}
\end{gather}
where $u_i$ is the $i^{\textrm{th}}$ velocity component ($i=1,2,3$), $E = p/(\gamma - 1) + 0.5 \rho u_i u_i$ is the total energy, and $\Theta = p / ( R\rho)$ is the temperature, with gas constant $R=287\,\mathrm{J / (kg\cdot K)}$. Thermal conductivity is given by $\kappa=\mu C_p /Pr$. The dynamic shear viscosity is modeled as $\mu=\mu_e (\Theta/\Theta_e)^{0.8497}$, while the Prandtl number $Pr=0.72$ is assumed constant, consistent with ideal gas behavior \citep{touloukian1975_thermophysical}. The viscous stress tensor $\tau_{ij}$ is given by
\begin{equation}
    \tau_{ij} = \mu \left( \frac{\partial u_i}{\partial x_j} + \frac{\partial u_j}{\partial x_i} \right) + \left( \mu_b-\frac{2}{3}\mu \right) \delta_{ij} \frac{\partial u_k}{\partial x_k}.
    \label{eq:NS4}
\end{equation}

The governing equations (\ref{eq:NS1}-\ref{eq:NS4}) can be expressed in compact form as $\boldsymbol{q} = \mathcal{N}(\boldsymbol{c})$, where $\boldsymbol{q} = [\rho, \rho u_1, \rho u_2, \rho u_3, E]^{\top}$ is the state vector, $\mathcal{N}$ is the Navier-Stokes operator, and $\boldsymbol{c}$ is the vector of control parameters which we must accurately estimate in order to reproduce the experimental data.   In the present study, $\boldsymbol{c}$ will be comprised of the oncoming boundary-layer disturbance spectra, at the inflow to the computational domain.  

\begin{figure}
    \centering
    \includegraphics[width=\textwidth]{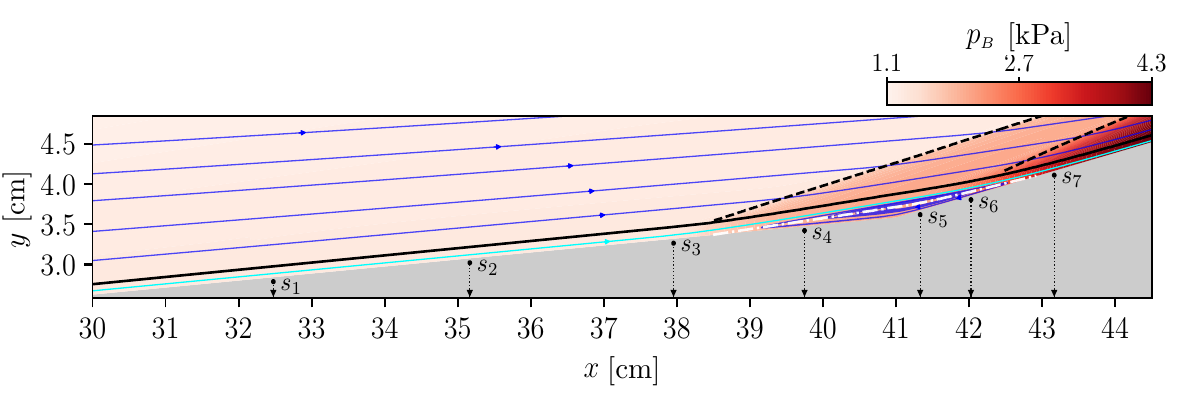}
    \caption{Contours of pressure in the axisymmetric undisturbed flow, $\boldsymbol{q}_{\scriptscriptstyle B}$. ({\protect\solidLine{0}{0}{0}}) The  boundary-layer thickness $\delta_{\scriptscriptstyle 99}$; ({\protect\dashedLine{0}{0}{0}}) separation and reattachment shocks; ({\protect\solidLine{0}{0}{1}}) velocity streamlines; ({\protect\solidLine{0}{1}{1}}) sonic line; ({\protect\dashedDottedLine{0}{0}{0}}, white) recirculation bubble identified by $u_{\xi,{\scriptscriptstyle B}}\!=\!0$; ($s_{1}$-$s_{7}$) sensor locations.}
    \label{fig:fig02_baseflow}
    \vspace{-6pt}
\end{figure}

The start of the computational domain on the cone surface is at $x_0 \! =  \! 29.86\,\mathrm{cm}$ ($\xi_{0} \! = \! 30.00\,\mathrm{cm}$), and the vertical extent is from the cone surface $\eta_0$ to $\eta_{max}\! = \! 2.55\,\mathrm{cm}$. The azimuthal domain size is $36^{\circ}$, which can accommodate linearly unstable three-dimensional modes and, as will be shown, is sufficiently large to capture the steady streaks in the assimilated flow.
The domain consists of a swept volume, where the inlet surface is translated along the cone-flare ($\xi$-axis) to the outlet.  
The downstream edge of the computational domain, on the cone surface, is $x_{max}\! = \!44.56\,\mathrm{cm}$. 
On the cone surface, no-slip and isothermal conditions are imposed, with the wall temperature fixed at $\Theta_{w} \! = \! 300~K$ as in the experiment.  
Periodic boundary conditions are applied in the azimuthal direction. 
A sponge region is included at the top ($\Delta \eta_{sponge} \! = \! 6.54\,\mathrm{mm}$) and outlet ($\Delta \xi_{sponge} \! = \! 0.8\,\mathrm{mm}$), where a forcing term dampens the fluctuations and drives the flow towards a reference laminar solution $\boldsymbol{q}_{\scriptscriptstyle B}$. The laminar state is shown in figure \ref{fig:fig02_baseflow}, and exhibits key features including the recirculation bubble and the separation and reattachment shocks.

\begin{figure}
    \centering
    \includegraphics[width=\textwidth]{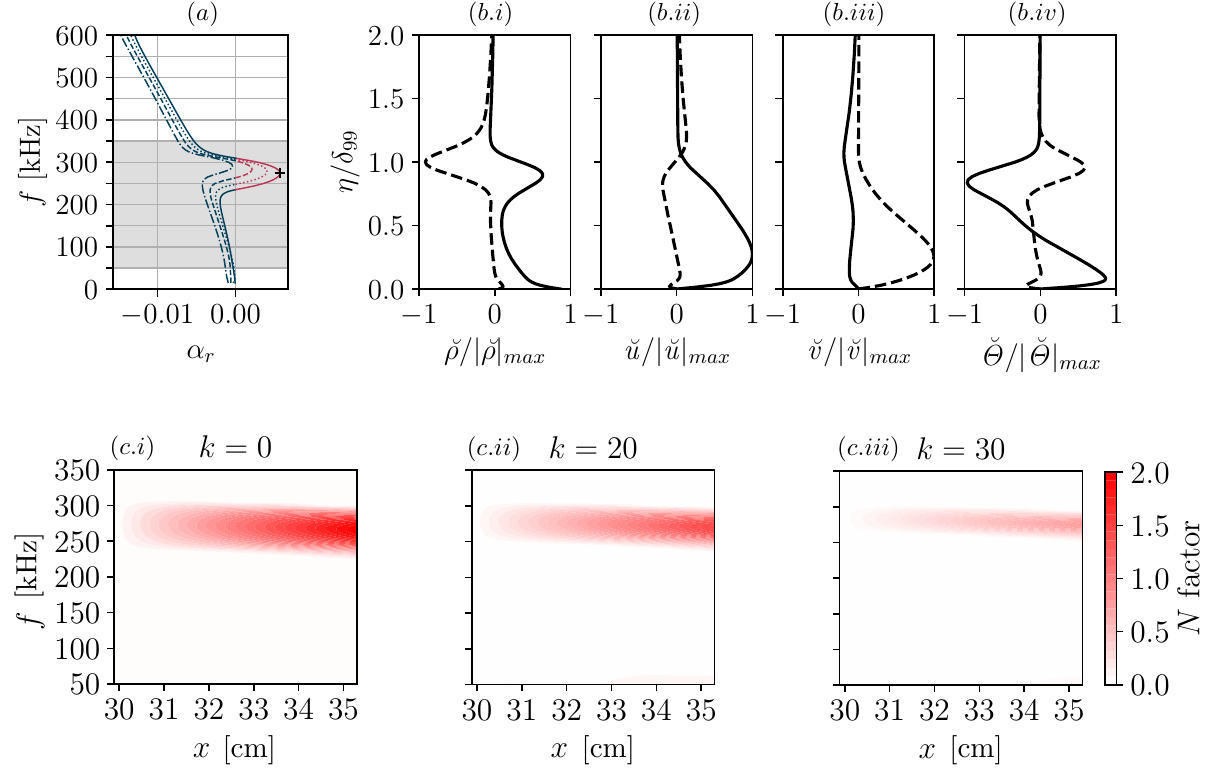}
    \caption{%
    (\textit{a}) Spatial growth rate $\alpha_r$ of the most unstable eigenfunction $\breve{\boldsymbol{q}}$ at given frequency-wavenumber pair ($f,\,k$), obtained from the linear stability analysis of the laminar axisymmetric flow $\boldsymbol{q}_{\scriptscriptstyle B}$ at the inflow. The largest value is indicated by the symbol ({\protect\colorplus{0}{0}{0}}), positive and negative values are distinguished by different colors; ({\protect\solidLine{0}{0}{0}}) $k=0$, ({\protect\dottedLine{0}{0}{0}}) $k=20$, ({\protect\dashedLine{0}{0}{0}}) $k=30$, ({\protect\dashedDottedLine{0}{0}{0}}) $k=40$.
    (\textit{b}) Wall-normal profiles of the most unstable mode $\breve{\boldsymbol{q}}_{n,m}$ marked in (\textit{a}); ({\protect\dashedLine{0}{0}{0}}) imaginary part, ({\protect\solidLine{0}{0}{0}}) real part.
    (\textit{c}) $N$-factor from linearized Navier-Stokes simulations about the laminar flow $\boldsymbol{q}_{\scriptscriptstyle B}$.
    }
    \label{fig:fig02_LSTmodes}
   \vspace{-6pt}
\end{figure}

In order to reproduce the experimental measurements, disturbances are introduced at the inlet plane, $\boldsymbol{x} = \boldsymbol{x}_{in}$, superposed onto the laminar base flow, $\boldsymbol{q} = \boldsymbol{q}_{\scriptscriptstyle B}(\boldsymbol{x}_{in})  +  \boldsymbol{q}^{\prime}_{\scriptscriptstyle \mathcal{L}}(\boldsymbol{x}_{in},t)$.  Subscript $\mathcal{L}$ indicates that these disturbances are computed using linear theory. Specifically, the disturbances are harmonic in time with frequency $f_n$ and in the span with wavenumber $k_m$; for each $\left(f_n,k_m\right)$ pair, the wall-normal profile corresponds to the most unstable discrete eigenfunction $\boldsymbol{\breve{q}}_{n,m}$ of the spatial linear-stability operator:
\begin{equation}\label{eq:disturbances}
    \boldsymbol{q}^{\prime}_{\scriptscriptstyle \mathcal{L}}(\boldsymbol{x}_{in}, t) = 
    \sum_{n,m} \frac{c_{n,m}}{\sqrt{2}} \mathfrak{Re}\! \left[ {\boldsymbol{\breve{q}}_{n,m}}
    e^{i (k_{m} \vartheta - 2\pi f_{n} t + \psi_{n,m}) }
     \right].
\end{equation}
Since equation~(\ref{eq:disturbances}) is evaluated at the inflow, the streamwise exponential $e^{i \alpha_{m,n} \xi_{in}}$ is constant and is absorbed in the amplitude and phase for every mode.
Parameterization of the inflow disturbances using the eigenfunctions $\boldsymbol{\breve{q}}_{n,m}$ from linear-stability theory is motivated by the prominence of the associated frequencies in the experimental wall-pressure spectra at the upstream-most sensors, as will be seen in \S\ref{sec:da_observations}.
The eigenfunctions are normalized to unit energy,
\begin{equation}
    \frac{1}{2\rho_{e}U_{e}^2\delta_e} \int_{\eta_0}^{\eta_{max}} 
    \left[
    \rho_{\scriptscriptstyle B} |\breve{u}_{i}|^2 + 
    R\,\frac{\Theta_{\scriptscriptstyle B}}{\rho_{\scriptscriptstyle B}} |\, \breve{\rho} |^2 + 
    \frac{R}{\gamma-1} \frac{\rho_{\scriptscriptstyle B}}{\Theta_{\scriptscriptstyle B}} | \breve{\Theta} |^2
    \right]
    \ {\mathrm{d}}\eta = 1, \quad \quad i = 1,2,3.
\end{equation}
using the norm by \cite{chu1965_energy}, where $\delta_e$ is the Mangler-transformed Blasius lengthscale at the inflow.

The prescribed inflow frequencies are $f_n \in \{50,\, \allowbreak 55,\allowbreak \dots,\allowbreak 350\}\,\mathrm{kHz}$ with $n \! = \! 1,2,\dots,N_{f}$ and $N_{f} \! = \! 61$, while the prescribed azimuthal wavenumbers are $k_m \in \{0,20,30,40\}$ with $m \! = \! 1,\dots,N_k$ and $N_k \! = \! 4$. These ranges were identified based on spectral analysis of the measurements (\S\ref{sec:da_observations}), and encompass Mack's first- and second-mode at the inlet plane. 
In figure \ref{fig:fig02_LSTmodes}($a$), 
we plot the local spatial growth rates $\alpha_r$ of these discrete modes, over an expanded range of frequencies. The eigenfunction associated with the peak growth rate is then reported in figure \ref{fig:fig02_LSTmodes}($b$).  As expected for modes in the discrete branch, $\breve{\boldsymbol{q}}_{n,m}(\eta)$ is contained within the boundary layer. 
Using the linearized Navier-Stokes equations, the inflow modes were evolved along the early portion of the cone and their $N$-factors are reported in figure \ref{fig:fig02_LSTmodes}($c$).  The results demonstrate that the planar and oblique waves with frequencies $f_n \in [225,310]\,\mathrm{kHz}$ are the linearly most amplified downstream. The data assimilation then exploits the amplification and nonlinear interactions among the inflow modes to match the experimental measurements, for example for the generation of higher harmonics of the inlet frequencies or the formation of streaky structures \citep{novikov2016_wavepackets}.
The linearly stable lower frequencies are also included at the inlet due to their relevance in the separated region, which can destabilize these waves.
The phases $\psi_{n,m}$ are sampled from independent uniform distributions, over the range $[-\pi, \pi)$.   The modal amplitudes constitute the control vector $\boldsymbol{c} = [\, \dots,\, c_{n,m}, \, \dots \,]^{\top}$ with dimension $N_{c} = 244$, which we will optimize during the data assimilation procedure in order to reproduce available measurements.

Similar to the compact notation for the nonlinear Navier-Stokes equations $\boldsymbol{q} \! = \! \mathcal{N}(\boldsymbol{c})$, we introduce the linearized counterpart $\boldsymbol{q}_{\scriptscriptstyle \mathcal{L}}^{\prime} \! = \! \mathcal{L}_{\boldsymbol{q}_{\scriptscriptstyle B}}(\boldsymbol{c})$.  The notation $\mathcal{L}_{\boldsymbol{q}_{\scriptscriptstyle B}}$ for the linearized Navier-Stokes operator reflects its dependence on the base state $\boldsymbol{q}_{\scriptscriptstyle B}$.  The solutions of both the nonlinear and linear systems of equations are performed in curvilinear coordinates within the domain shown in figure~\ref{fig:fig01} (see \cite{vishnampet2015_practical} for details).  Time integration is based on a fourth-order Runge-Kutta scheme and fourth-order centered finite differences are adopted for the spatial derivatives. At the boundaries, second-order finite differences with biased stencils are applied. Second derivatives are computed by applying the discretized first-derivative operator twice \citep{mattsson2004_summation}.

\begin{table}
    \centering
    \begin{tabular}{lcccccc}
     & & \multicolumn{2}{l}{$(~ \, x_{in}~ ,~~ \, x_{f}~,~x_{max}\, )\, \mathrm{[cm]} $}& \multicolumn{2}{c}{$L_{\eta_{in}}\, \mathrm{[cm]}$}& $L_{\vartheta}\, \mathrm{[deg.]}$\\[0.4em]
     \multicolumn{2}{l}{Domain size} & \multicolumn{2}{l}{$(29.86,\, \,41.00,\, \, 44.56)$}& \multicolumn{2}{c}{$2.55$}& $36^{\circ}$\\[0.4em] \hline \\[-0.4em]
     &
     $N_{\xi}\! \times \! N_{\eta} \! \times \! N_{\vartheta}$&
     $\Delta \xi_{in}\, \mathrm{[cm]}$ & $\Delta \xi_{out}\, \mathrm{[cm]}$ & 
     $\Delta \eta_{w,in}\, \mathrm{[cm]}$ & $\Delta \eta_{w,out}\, \mathrm{[cm]}$ & 
     $\Delta \vartheta\, \mathrm{[rad]}$ \\[0.5em]
     Grid G1&
     $1053 \! \times \! 201 \! \times \! 64$&
     $1.40 \! \times \! 10^{-2}$ & $1.44 \! \times \! 10^{-2}$&
     $6.78 \! \times \! 10^{-4}$ & $6.57 \! \times \! 10^{-4}$&
     $9.82 \! \times \! 10^{-3}$ \\
     Grid G2 &
     $1728\! \times \! 230 \! \times \! 108$ &
     $1.40 \! \times \! 10^{-2}$ & $3.33 \! \times \! 10^{-3}$ & $5.91 \! \times \! 10^{-4}$ & $1.20 \! \times \! 10^{-3}$ &
     $5.82 \! \times \! 10^{-3}$
    \end{tabular}
    \caption{Domain sizes and grid resolutions. Subscripts `$in$', `$out$', `$w$' indicate grid cells at the inlet, outlet, and the wall. The $x$-axis values are on the surface $\eta_0$, and subscript `$f$' denotes the cone-flare corner point.}
    \label{tab:grid}
\end{table}

Two grids will be used to simulate the flow over the cone-flare geometry, and the associated parameters are provided in table~\ref{tab:grid}. The first grid (G1) is designed to resolve the flow from the inlet to the third sensor, and will be adopted in an initial assimilation that considers the first two sensors ($s_1,\, s_2$) only.  The outcome of this assimilation will enable us to examine the accuracy of predicting the downstream flow from upstream measurements.  
The second grid (G2) is finer than G1 downstream of the second sensor, and is designed to fully capture all flow features throughout the domain. This grid will be used to assimilate the measurements from all seven sensors.  
Both grids feature uniform spacing along the azimuthal direction $\vartheta$ and employ the grid-stretching technique used in \cite{pruett1995_spatial} in the wall-normal direction, increasing resolution near the wall and across the boundary layer edge. 
Grid G1 has a fixed $\Delta x$, while grid G2 undergoes hyperbolic tangent refinement in the streamwise direction.

For each grid, simulations are performed for a sufficiently long duration to clear transient effects prior to acquiring data.
The simulation time step is set to $\Delta t = 1/(175\,\mathrm{MHz})$, which ensures that the CFL remains consistently below $0.35$. Data are sampled from the simulation at $17.5\,\mathrm{MHz}$, which is approximately three orders of magnitude higher than the highest frequency at the inflow ($350\,\mathrm{kHz}$) and in the experimental spectra described in \S\ref{sec:da_observations}. For the assimilation, observation data are collected for $0.2\,\mathrm{ms}$, and therefore the minimum resolved frequency is $5\,\mathrm{kHz}$ which is one-tenth of the lowest frequency at the inflow ($50\,\mathrm{kHz}$). For the analysis of the final assimilated state, data are collected for $1\,\mathrm{ms}$ (or $1\,\mathrm{kHz}$ resolution).

Mean quantities are denoted $\overline{\bullet}$ and are computed by averaging in time and the azimuthal direction. When averaging is performed in only one coordinate, it is marked as $\overline{\bullet}^{\, t}$ or $\overline{\bullet}^{\, \vartheta}$. 
Fourier transformed quantities are denoted by $\widehat{\bullet}$ for transforms in time and by $\widehat{\widehat{\bullet}}$ for transforms in both time and azimuthal direction.
Filtered quantities are denoted $\langle \bullet \rangle_{f}$ where the subscript denotes the retained frequencies.

\section{Data assimilation}
\label{sec:da}
%
\subsection{Experimental measurements and model observations}
\label{sec:da_observations}
\begin{figure}
    \centering
    \includegraphics[width=\textwidth]{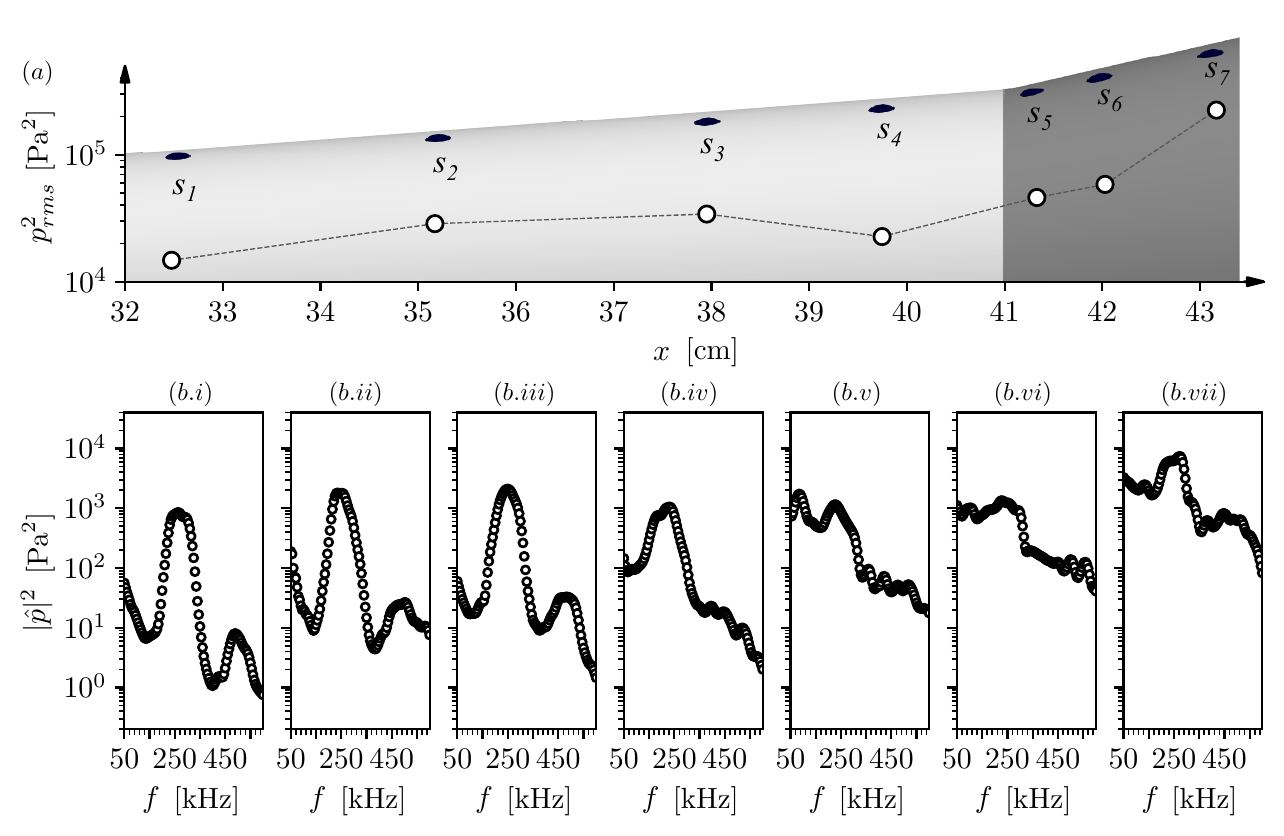}
    \caption{Experimental measurements. (\textit{a}) Wall-pressure intensity and (\textit{b}) frequency spectra at the sensors locations.}
    \label{fig:fig03}
    \vspace*{-4pt}
\end{figure}
The experimental measurements used in the data assimilation are presented in figure~\ref{fig:fig03}. These data consist of the wall-pressure spectra $|\widehat{p}|^{2}(f)$ and intensity 
recorded by the seven ($N_{s}\!=\!7$) PCB sensors $\{s_1, s_2, s_3, s_4, s_5, s_6, s_7 \}$ described in \S\ref{sec:config_exp}, during $6\,\mathrm{ms}$ after an initial transient is discarded. The intensity was evaluated from the spectra, $p^2_{rms} = \sum_{j=1}^{M_{f}} |\widehat{p}(f_{j})|^2$, using the frequencies $f \in \{50,55,\dots, 600\}\,\mathrm{kHz}$ and $M_f = 111$. In the experiments, the wall pressure is recorded at a sampling rate of $2\,\mathrm{MHz}$ and a $600\,\mathrm{kHz}$ low-pass filter removes aliasing effects (see \cite{butler2021_secondmode} for details). The spectra are evaluated with the Welch method, which involves averaging an ensemble of realizations. Each realization is generated by extracting a $0.2\,\mathrm{ms}$ segment of the time series, applying a Hann window to minimize spectral leakage, and then performing a Fourier transform. Consecutive segments within the full $6\,\mathrm{ms}$ recording period have $50\%$ overlap, which yields fifty-nine segments and a relative standard deviation of $1/\sqrt{59}$. The noise floor of the PCB sensors, estimated from a $6~\mathrm{ms}$ recording acquired prior to flow arrival, is approximately two orders of magnitude lower than the flow induced spectra across the relevant frequency range and is therefore neglected \citep{butler2022_coneflare}.

The intensity in figure~\ref{fig:fig03} rises from the first through the seventh sensor, with the exception of sensor four where there is an appreciable reduction in wall-pressure fluctuations. 
The spectra in figure~\ref{fig:fig03} reveal that, for the first three sensors, the dynamics are dominated by the frequencies in the range $[240,305]\,\mathrm{kHz}$, known to be unstable from linear theory. 
At these locations, a secondary peak amplifies at the harmonics of the dominant frequencies, consistent with the outcome of nonlinear effects. 
At sensor four, the amplitudes of both the dominant frequencies and their higher harmonics are reduced. A qualitative change is observed at sensor five, where the amplitudes of frequencies below $150\,\mathrm{kHz}$ become large.  
At sensor seven, the frequencies $f \in [240, 305]\,\mathrm{kHz}$ increase in amplitude again, and their harmonics reemerge.
The changes in the spectra from sensor $s_4$ to $s_5$, specifically the amplification of low-frequency disturbances, is consistent with reports of low-frequency disturbance amplification within recirculation regions in both previous computations \citep{paredes2022_cone} and experiments \citep{butler2022_coneflare}. The measurement data, however, lack details on the three-dimensional flow structures and the role of the compression shock in the disturbance dynamics. While these details are present in simulations, only data assimilation can ensure that the simulated phenomena reproduce the measurements quantitatively. 

For the purpose of data assimilation, the experimental wall-pressure spectra and intensities from the seven sensors are concatenated into two vectors, 
\begin{align}
    \boldsymbol{m}_{\scriptscriptstyle S} &= 
       \left[\, \dots,\,
          \log_{10}|\widehat{p}(s_i,f_{j})|^2,
          \, \dots \, \right]^{\top}\label{eq:mSexp},\\[0.5em]
    \boldsymbol{m}_{\scriptscriptstyle I} &= 
        [\, \dots,\,
          \sum_{j}|\widehat{p}(s_i,f_{j})|^2,
        \, \dots\, ]^{\top}\label{eq:mIexp},
\end{align}
where $i = 1,\, \dots,\, N_{s}$ and $j=1,\, \dots,\, M_{f}$. The dimensions of the measurements are $\boldsymbol{m}_{\scriptscriptstyle S} \in \mathbb{R}^{M_{\scriptscriptstyle S} \times 1 } $ with $M_{\scriptscriptstyle S} = M_f N_s = 777$ and $\boldsymbol{m}_{\scriptscriptstyle I} \in \mathbb{R}^{M_{\scriptscriptstyle I} \times 1 }$ with $M_{\scriptscriptstyle I} = N_s = 7$.  These experimental measurements will be compared to their numerical counterpart from the simulations.

\subsection{Ensemble-variational (EnVar) algorithm}
\label{sec:da_algorithm}%
The goal of data assimilation is to identify the unknown amplitudes $\boldsymbol{c}$, referred to as the control vector, which in the simulations reproduce the experimental measurements $\boldsymbol{m}_{\scriptscriptstyle S}$ and $\boldsymbol{m}_{\scriptscriptstyle I}$.  This problem is formulated as the minimization of a scalar cost function $\mathcal{J}(\boldsymbol{c})$, which is defined in terms of the disparity between the available measurements and their numerical estimation, 
\begin{equation}
    \mathcal{J}(\boldsymbol{c}) = \underbrace{\frac{1}{2}|\!| \boldsymbol{m}_{\scriptscriptstyle S} - \mathcal{M}_{\scriptscriptstyle S}(\boldsymbol{c}) |\!|_{\boldsymbol{\Sigma}_{\scriptscriptstyle S}^{-1}}^{2}}_{\mathcal{J}_{\scriptscriptstyle S}(\boldsymbol{c})} + 
    \underbrace{\frac{1}{2}|\!| \boldsymbol{m}_{\scriptscriptstyle I} - \mathcal{M}_{\scriptscriptstyle I}(\boldsymbol{c}) |\!|_{\boldsymbol{\Sigma}_{\scriptscriptstyle I}^{-1}}^{2}}_{\mathcal{J}_{\scriptscriptstyle I}(\boldsymbol{c})} +
    \underbrace{\frac{1}{2}|\!| \boldsymbol{c} - \boldsymbol{c}_{i}|\!|_{\boldsymbol{\Sigma}_{c,i}^{-1}}^{2}}_{\mathcal{J}_{\scriptscriptstyle P}(\boldsymbol{c})}, 
    \label{eq:costDA}
\end{equation}
where $|\!| \bullet |\!|_{\boldsymbol{\Sigma}^{-1}}^{2} = \bullet^\top \boldsymbol{\Sigma}^{-1} \bullet$.  
In the above expression, $\mathcal{M}_{\scriptscriptstyle S}(\boldsymbol{c})$ and $\mathcal{M}_{\scriptscriptstyle I}(\boldsymbol{c})$ are the computational estimates of the wall-pressure spectra and intensities, respectively, resulting from the amplitudes $\boldsymbol{c}$.  To ensure consistency, the wall-pressure signals are recorded at the same sensor locations as in the experiments (\S\ref{sec:config_exp}) during $0.2\,\mathrm{ms}$, and the spectra are evaluated at the same frequency resolution of $5\,\mathrm{kHz}$.  Since the experimental data do not provide information in the azimuthal direction, the frequency spectra in the computations are averaged over all spanwise locations.  
The discrepancies in reproducing the spectra and intensity are weighted, respectively, by the covariance matrices $\boldsymbol{\Sigma}_{\scriptscriptstyle S} \in \mathbb{R}^{M_{\scriptscriptstyle S} \times M_{\scriptscriptstyle S}}$ and $\boldsymbol{\Sigma}_{\scriptscriptstyle I} \in \mathbb{R}^{M_{\scriptscriptstyle I} \times M_{\scriptscriptstyle I}}$, which model
the uncertainty in the experimental data.  
These matrices are assumed to be diagonal, with entries $\Sigma_{S,ii} = M_S (0.01\, m_{S,i})^2$ and $\Sigma_{I,ii} = M_I (0.01\, m_{I,i})^2$; the normalization factors $M_{S,I}$ which are absorbed in $\Sigma_{S,I}$ are such that $\mathcal{J}_S$ and $\mathcal{J}_I$ measure average squared relative misfits, to prevent $\mathcal{J}_S$ from dominating the cost function solely because it contains more entries.
The last term in the cost function is the departure of the optimal control vector from the prior estimate $\boldsymbol{c}_{i}$, and the covariance $\boldsymbol{\Sigma}_{c}  \in \mathbb{R}^{N_{\scriptscriptstyle c} \times N_{\scriptscriptstyle c}}$ reflects the uncertainty in that prior.  

The definition of the cost function is important, and the above form has proven effective in the assimilation of various hypersonic-flow datasets \citep{buchta2021_observation,buchta2022_cone}. The first term, $\mathcal{J}_{\scriptscriptstyle S}$, is based on the logarithm of the wall-pressure spectra. Since the spectra span several orders of magnitude, the logarithmic scale ensures that each frequency contributes similarly to the cost, preventing the optimization from focusing solely on dominant frequencies. The second term, $\mathcal{J}_{\scriptscriptstyle I}$, incorporates the intensity to capture the overall spectral peak at each sensor. 

The minimization of $\mathcal{J}(c)$ is performed using an ensemble variational (EnVar) approach. Starting with an estimate of the control vector, the local gradient and Hessian are approximated using an ensemble of perturbations to the control vector and their associated measurements. The estimated control vector is then updated in the direction of steepest descent, and the process is repeated until convergence.  

Mathematically, we start with the estimate $\boldsymbol{c}_i$.  We additionally introduce an ensemble of control vectors $\boldsymbol{c}_i^{(j)}$ ($j\!=\! 1,\dots, N_{ens}$) that have $\boldsymbol{c}_i$ as their mean and a covariance $\boldsymbol{\Sigma}_{c}$.  The optimal control vector is then expressed as the weighted superposition, 
\begin{equation}
    \boldsymbol{c} = \boldsymbol{c}_i + 
    \mathsfbi{P}\boldsymbol{w}, 
    \label{eq:ensToControlVec}
\end{equation}
where $\mathsfbi{P} = [\dots,\,  \boldsymbol{c}_i^{(j)} - \boldsymbol{c}_i,\, \dots] \in \mathbb{R}^{N_{c} \times N_{ens}}$ and $\boldsymbol{w} = [\dots,w_j,\dots]^{\top} \in \mathbb{R}^{N_{ens} \times 1}$ are the optimal weights. 
In the form~(\ref{eq:ensToControlVec}), the control vector becomes the weights $\boldsymbol{w}$. The cost function is then approximated by the quadratic form, 
\begin{equation}
    \widetilde{\mathcal{J}}(\boldsymbol{w}) \! = \! \frac{1}{2} |\!| \boldsymbol{m}_{\scriptscriptstyle S}  -  \mathcal{M}_{\scriptscriptstyle S}(\boldsymbol{c}_i)  -\mathsfbi{H}_{ \scriptscriptstyle S} \boldsymbol{w}|\!|_{\! {\boldsymbol{\Sigma}}_{\scriptscriptstyle S}^{-1}}^{2} + 
    \frac{1}{2}|\!| \boldsymbol{m}_{\scriptscriptstyle I}  - \mathcal{M}_{\scriptscriptstyle I}(\boldsymbol{c}_i) -\mathsfbi{H}_{ \scriptscriptstyle I} \boldsymbol{w}|\!|_{\! \boldsymbol{\Sigma}_{\scriptscriptstyle I}^{-1}}^{2} +
    \frac{1}{2}|\!| \mathsfbi{P}\boldsymbol{w}|\!|_{\! \boldsymbol{\Sigma}_{\! c}^{-1}}^{2},
    \label{eq:costEnVarapprox}
\end{equation}
where $\mathcal{M}(\boldsymbol{c}_i)$ are the observations associated with the prior estimate $\boldsymbol{c}_i$, and $\mathsfbi{H} = [\, \dots,\, \mathcal{M}(\boldsymbol{c}_i^{(j)}) - \mathcal{M}(\boldsymbol{c}_i) ,\, \dots \,]$ is the observation perturbation matrix of the ensemble.  
For optimality, we assume that the gradient of $\widetilde{\mathcal{J}}(\boldsymbol{w})$ vanishes and solve for the optimal weights, 
\begin{equation}
\label{eq:wstar}
    \boldsymbol{w} = \left( \frac{\partial^{2} \widetilde{\mathcal{J}}}{\partial \boldsymbol{w}\partial \boldsymbol{w}^{\top}} \right)^{-1} 
    \left[ 
     \mathsfbi{H}_{\scriptscriptstyle S}^{\top}\boldsymbol{\Sigma}_{\scriptscriptstyle S}^{-1}(\boldsymbol{m}_{\scriptscriptstyle S} - \mathcal{M}_{\scriptscriptstyle S}(\boldsymbol{c}_i)) + 
      \mathsfbi{H}_{\scriptscriptstyle I}^{\top}\boldsymbol{\Sigma}_{\scriptscriptstyle I}^{-1}(\boldsymbol{m}_{\scriptscriptstyle I} - \mathcal{M}_{\scriptscriptstyle I}(\boldsymbol{c}_i))
    \right],
\end{equation}
where 
\begin{equation}
\frac{\partial^{2} \widetilde{\mathcal{J}}}{\partial \boldsymbol{w}\partial \boldsymbol{w}^{\top}} = 
    \mathsfbi{H}_{\scriptscriptstyle S}^{\top}\boldsymbol{\Sigma}_{\scriptscriptstyle S}^{-1} \mathsfbi{H}_{\scriptscriptstyle S}^{}+ 
    \mathsfbi{H}_{\scriptscriptstyle I}^{\top}\boldsymbol{\Sigma}_{\scriptscriptstyle I}^{-1}\mathsfbi{H}_{\scriptscriptstyle I}^{} +
\mathsfbi{P}^{\top}\boldsymbol{\Sigma}_{c}^{-1}\mathsfbi{P}^{}.
\end{equation}
The optimal weights (\ref{eq:wstar}) from the quadratic approximation $\mathcal{\widetilde{J}}$ are not necessarily optimal for the original $\mathcal{J}$.  A line-search along the direction $\alpha \boldsymbol{w}$ is performed using Jaratt's method to identify the minimizer of $\mathcal{J}(\boldsymbol{c}_{i} + \alpha\mathsfbi{P}\boldsymbol{w})$.  
Once the optimal $\alpha$ is found, the control vector is updated according to,
\begin{equation}
    \boldsymbol{c}_{i+1} = \boldsymbol{c}_{i} + \alpha \mathsfbi{P}\boldsymbol{w}.
\end{equation}
After each iteration of this procedure, the uncertainty in the estimated control vector is reduced, and therefore the ensemble covariance is updated to reflect this change, 
\begin{equation}
    \boldsymbol{\Sigma}_{c,i+1} = \frac{1}{N_{ens}-1} \mathsfbi{P}_{i+1}^{} \mathsfbi{P}_{i+1}^{\top},  
    \quad \textrm{where} \quad     
    \mathsfbi{P}_{i+1} = \sqrt{N_{ens}-1} \mathsfbi{P}_{i} \left( \frac{\partial^{2} \widetilde{\mathcal{J}}}{\partial \boldsymbol{w}\partial \boldsymbol{w}^{\top}} \right)^{-1/2}\mathsfbi{U},
    \label{eq:covCupdate}
\end{equation}
and 
$\mathsfbi{U} \in \mathbb{R}^{N_{ens} \times N_{ens}}$ is a random, mean preserving, unitary matrix.  The entire procedure is repeated until the cost function is reduced to a desired level where the simulation reproduces the measurements with sufficient accuracy.

\emph{Generation of initial estimate and ensemble:}
To initialize the algorithm, a first estimate of the control vector is needed, and is computed using the linearized Navier-Stokes equations.  Since the assumption of linear dynamics becomes progressively more inaccurate as instabilities amplify with downstream distance, only the first sensor data are used in the definition of the cost function, 
\begin{equation}
    \mathcal{J}_{\scriptscriptstyle \mathcal{L}}(\boldsymbol{c}) = \frac{1}{2} |\!| \boldsymbol{m}_{s_{1}}  - \mathsfbi{L} \boldsymbol{c} |\!|^2 + \frac{\gamma}{2} |\!| \boldsymbol{c} |\!|^2,
    \label{eq:linearDA}
\end{equation}
where $\boldsymbol{m}_{s_{1}} = [\ \dots,\, |\widehat{p}(s_1,f)|^2,\, \dots \ ]^{\top} \in \mathbb{R}^{M_f \times 1}$ are the pressure spectra at the first sensor $s_{1}$. The matrix $\mathsfbi{L} = [ \ \dots, \, \boldsymbol{l}_{n,m},\, \dots \ ] \in \mathbb{R}^{M_f \times N_{f}N_k}$ is comprised of columns $\boldsymbol{l}_{n,m} = [ 0,\ \dots, 0, \, |\widehat{p}_{\scriptscriptstyle \mathcal{L}}(f_n,k_m)|^2,\, 0,\ \dots, 0 ]^{\top} \in \mathbb{R}^{M_f \times 1}$, each corresponding to the measurements from a unit-amplitude upstream instability wave with wavenumber $(n,m)$.  Multiple solutions are possible because the measurements do not distinguish two- and three-dimensional waves.  For this reason, we introduce the regularization parameter $\gamma$ which presents a tradeoff between the accuracy of reproducing the measurements and the total energy of the linear estimate of $\boldsymbol{c}$. The minimizer of (\ref{eq:linearDA}) is, 
\begin{equation}
    \widetilde{\boldsymbol{c}}_{0} = 
    (\mathsfbi{L}^{\top}\mathsfbi{L} + \gamma \mathsfbi{I})^{-1}\mathsfbi{L}^{\top}\boldsymbol{m}_{s_{1}}.
    \label{eq:linearDAguess}
\end{equation}

The covariance of the initial ensemble is defined by a Gaussian kernel function, ${\Sigma}_{c,ij} = (0.05\,\widetilde{c}_{0,i})^2\exp[ -(f_i - f_j)^2/\sigma_f^2 -(k_i - k_j)^2/\sigma_k^2]$, where $i,j=1,\dots,N_{c}$.  The correlation lengths $\sigma_f$ and $\sigma_k$ are optimized to ensure that the leading ten ($N_{ens}=10$) eigenvectors of the covariance matrix equally accurately represent any delta function in the space spanned by $\boldsymbol{c}$ \citep[see][]{mons2021_ensemble}.  The initial ensemble perturbations are then formed as, 
$\mathsfbi{P}_{0} = \sqrt{(N_{ens}-1)}\ \mathsfbi{{W}}\boldsymbol{{\Lambda}}^{{\scriptscriptstyle \, 0.5}}\mathsfbi{U}$
where the columns of $\mathsfbi{{W}}$ are the ten ($N_{ens}=10$) eigenvectors, and the diagonal matrix $\boldsymbol{\Lambda}$ has the associated eigenvalues, and the initial covariance matrix is $\boldsymbol{\Sigma}_{c,0} = \mathsfbi{P}_{0}^{}\mathsfbi{P}_{0}^{\top}/(N_{ens-1})$.
%
%
\subsection{Optimization protocol}
\label{sec:da_protocol}%
The assimilation will initially be performed using measurements from the first two sensors only.  During the iterative procedure, we will adopt grid G1 from table \ref{tab:grid} for computational efficiency, and the predicted control vector will be identified by a tilde ($\boldsymbol{{\widetilde{c}}}$).   Assuming that the predicted flow reproduces the data from these two sensors, we will examine whether it also reproduces the unassimilated measurements from the remaining downstream sensors.  For this step, we will use the finer grid G2.  In addition, the final prediction of ($\boldsymbol{{\widetilde{c}}}$) will be adopted as the initial guess for a new assimilation that considers all seven sensors to predict $\boldsymbol{c}$, and which adopts grid G2.  

In both the assimilation tasks, every iteration is comprised of $N_{ens}+1 = 11$ direct numerical simulations (DNS), which correspond to the current estimate of the control vector and the ten ensemble members.  In addition, two to four more DNS are conducted during the line search with Jaratt's method. 
The computational cost of the data assimilation scales with the number of ensemble members, the number of iterations, and the computational expense of each simulation. Each DNS using grid G1 requires only $8{,}200$ CPU-hours, while a DNS of grid G2 requires $30{,}000$ CPU-hours.

\section{Results}
\label{sec:results}

\subsection{Assimilation of the first two sensors data}
\label{sec:results_coarse}%
\begin{figure}
    \centering
    \includegraphics[width=\textwidth]{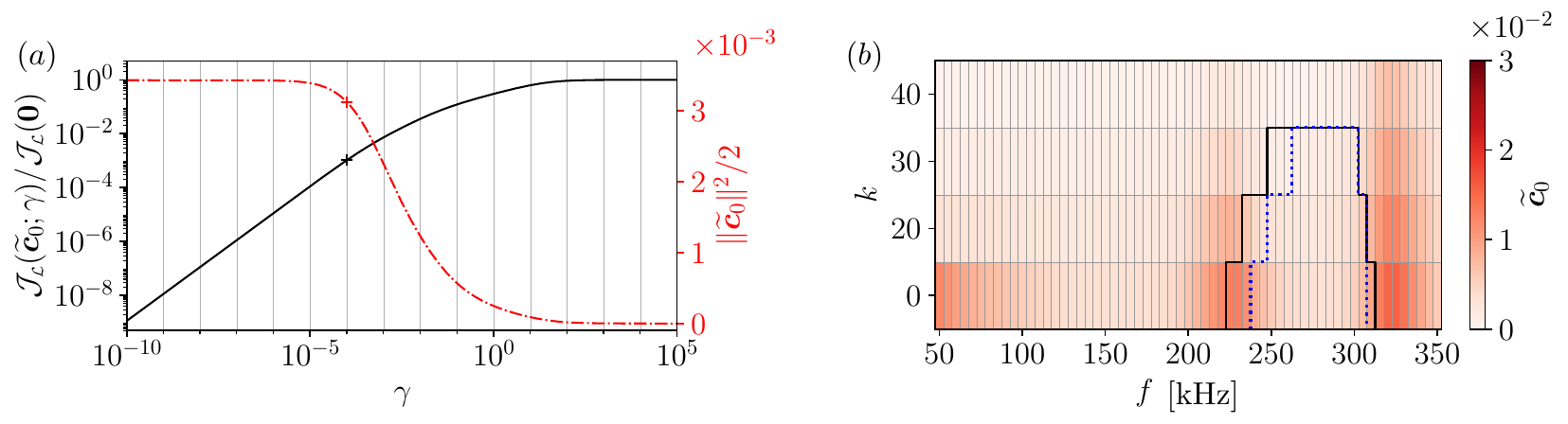}
    \caption{Initial estimate of the control vector. ($a$) Normalized linear cost function ({\protect\solidLine{0}{0}{0}}) and the energy of the inflow disturbance ({\protect\dashedDottedLine{1}{0}{0}}) plotted versus the regularization parameter, $\gamma$.  The adopted value of $\gamma$ is marked by a plus. ($b$) Initial estimate of the inflow disturbance spectra, computed using linear theory (\ref{eq:linearDAguess}) at the marked value of $\gamma$ in panel $(a)$. Lines mark linearly unstable modes at ({\protect\dottedLine{0}{0}{1}}) inflow and ({\protect\solidLine{0}{0}{0}}) according to the $N$-factor on the cone.
}
    \label{fig:fig04}
    \vspace*{-4pt}
\end{figure}
\begin{figure}
    \centering
    \includegraphics[width=\textwidth]{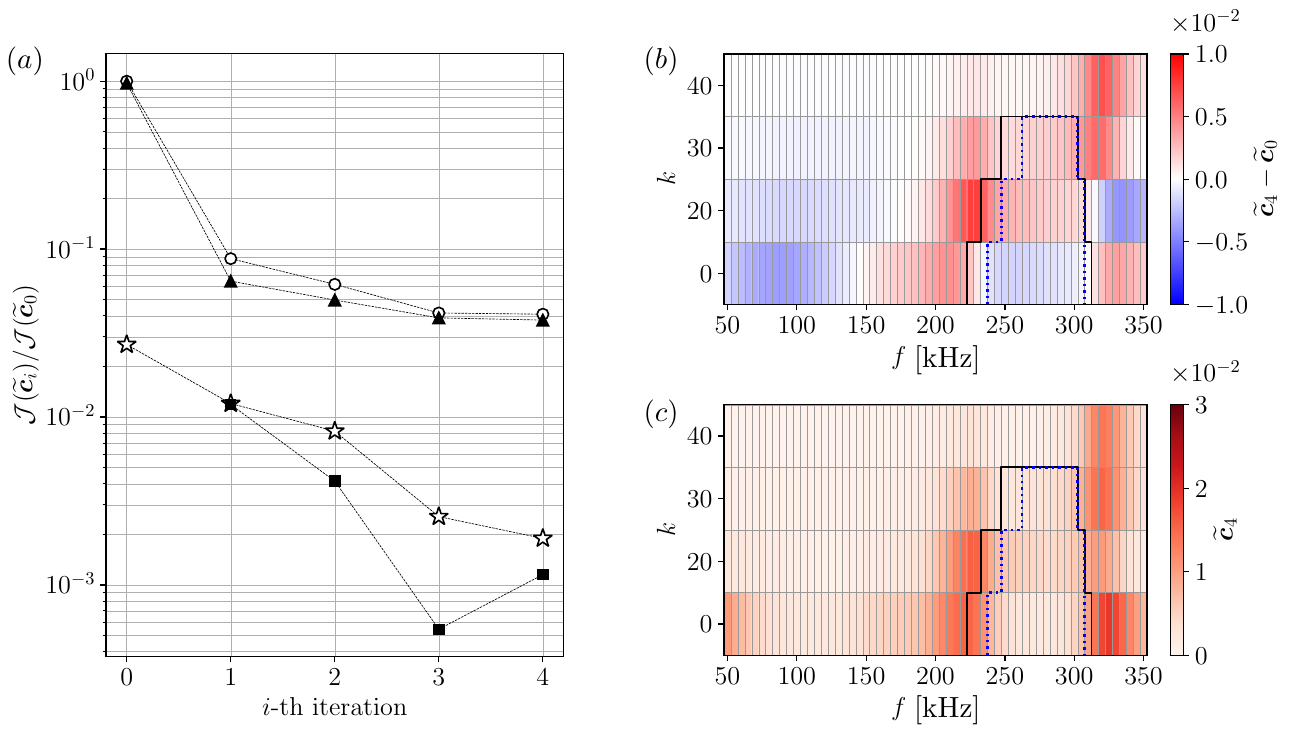}
    \caption{EnVar assimilation of the first two sensors, on grid G1. (\textit{a}) Terms in the cost function normalized by the initial total cost $\mathcal{J}(\boldsymbol{\widetilde{c}}_0)$.  
    ({\protect\colorcircle{1}{1}{1}{0}{0}{0}}) $\mathcal{J}$; ({\protect\colortriangle{0}{0}{0}{0}{0}{0}}) $\mathcal{J}_{\scriptscriptstyle S}$; ({\protect\colorstar{1}{1}{1}{0}{0}{0}}) $\mathcal{J}_{\scriptscriptstyle I}$; ({\protect\colorsquare{0}{0}{0}{0}{0}{0}}) $\mathcal{J}_{\scriptscriptstyle P}$. 
    (\textit{b}) Difference between the spectra of the final assimilated control vector and its initial estimate. (\textit{c}) Spectra of the final assimilated control vector. Lines in ($b$) and ($c$) mark linearly unstable modes at   ({\protect\dottedLine{0}{0}{1}}) inflow and ({\protect\solidLine{0}{0}{0}}) according to the $N$-factor on the cone.}
    \label{fig:fig05}
    \vspace*{-4pt}
\end{figure}
\begin{figure}
    \centering
    \includegraphics[width=.8\textwidth]{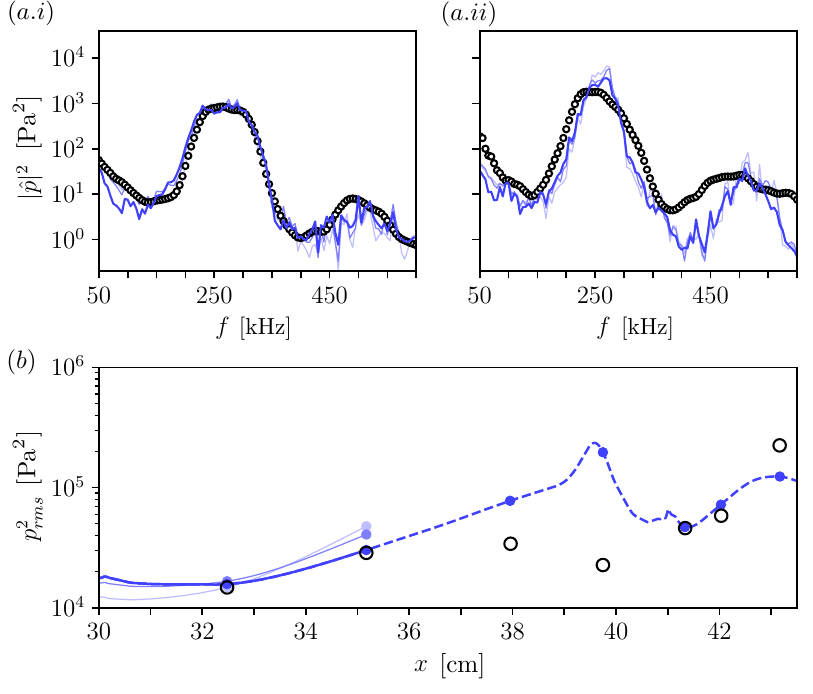}
    \caption{Wall-pressure spectra and intensity when assimilating the first two sensors. (\textit{a.i,a.ii}) Wall-pressure spectra at sensors $s_1$ and $s_2$. (\textit{b}) Wall-pressure intensity as a function of $x$. Black circles ({\protect\colorcircle{1}{1}{1}{0}{0}{0}}) indicate experimental measurements, solid lines ({\protect\solidLine{0}{0}{1}}) denote simulation results, and blue circles ({\protect\colorcircle{0}{0}{1}{0}{0}{1}}) mark intensities at sensor locations.  Light-to-dark blue represents EnVar iterations zero, one, and four.  The dashed line ({\protect\dashedLine{0}{0}{1}}) in (\textit{b}) shows prediction from $\boldsymbol{\widetilde{c}}_{4}$ using grid G2, and comparison to the measurements from sensors $s_3$ through $s_7$, which were not included in the assimilation. 
     }
    \label{fig:fig06}
    \vspace*{-4pt}
\end{figure}

Our starting point is an assimilation of the first two sensors.  The basic question is whether an accurate prediction of the early flow is sufficient to reproduce the downstream dynamics, including for example the onset of separation and its extent.  The concern here is not solely one of Lyapunov instability, where two infinitesimally close trajectories (the true flow and our assimilated state) of a chaotic system are bound to diverge in forward time, or downstream in a boundary layer.  In the present context, another effect arises that is specific to data assimilation. The observability of the first two sensors may not necessarily span all relevant upstream disturbances, and therefore the predicted flow may be missing important information that is relevant to the downstream dynamics.  Another related possibility is that the assimilated state is contaminated by disturbances that do not affect the upstream sensors, or in their null space, which lead to poor predictions of the downstream flow.  

The initial estimate $\boldsymbol{\widetilde{c}}_0$ is computed using equation~(\ref{eq:linearDAguess}), and is shown in figure~\ref{fig:fig04}. The regularization term $\gamma$ is selected based on a tradeoff between accuracy of the linear estimate and avoiding excessively large energy of the inflow disturbance $|\!| \boldsymbol{\widetilde{c}}_{0} |\!|^2 / 2$. Recall that the measurements are available at a single azimuthal location, and do not provide any information regarding the azimuthal variation.  This setup implies that, for a given frequency, disturbances with different amplitude combinations across azimuthal wave-numbers can reproduce the measurements, resulting in an infinite set of possible solutions. In this scenario, the regularization term helps select the solution with the smallest energy content from this infinite set.

The regularization parameter is chosen as $\gamma = 10^{-4}$, corresponding to the control vector $\boldsymbol{\widetilde{c}}_0$ shown in figure~\ref{fig:fig04}$b$. Within this control vector, the highlighted unstable range of frequencies does not exhibit the highest amplitudes.  These modes are the most efficiently amplified by the flow, and therefore do not require a large initial value to influence the downstream measurements.  The largest amplitudes are assigned to the stable modes which decay as they approach the first sensor. Planar waves, identified as the most unstable or least stable from linear theory, dominate the control vector.

Starting from the initial estimate $\boldsymbol{\widetilde{c}}_0$ and the associated covariance matrix $\boldsymbol{\widetilde{\Sigma}}_{c,0}$, four EnVar iterations are performed on grid G1, which reduce the normalized cost function by approximately one and a half orders of magnitude (figure~\ref{fig:fig05}). 
The total cost is dominated by the mismatch in frequency spectra $\mathcal{J}_{\scriptscriptstyle S}$ and the cost reduction is primarily due to an improvement in this component. Differences in overall intensity, captured by $\mathcal{J}_{\scriptscriptstyle I}$, also decrease as the spectral fit improves but remain less significant throughout. The prior term $\mathcal{J}_{\scriptscriptstyle P}$ stays subdominant and exhibits a sharp decrease to a minimum, suggesting that intermediate updates follow a steep descent. 
Compared to the initial estimate $\boldsymbol{\widetilde{c}}_0$, the converged solution $\boldsymbol{\widetilde{c}}_4$ (figure \ref{fig:fig05}$c$) exhibits increased amplitudes for both stable and unstable three-dimensional waves. However, the amplitudes of the unstable planar waves are reduced. These differences are shown explicitly by plotting the difference $\boldsymbol{\widetilde{c}}_4 - \boldsymbol{\widetilde{c}}_0$ in figure \ref{fig:fig05}$b$.

The predicted wall-pressure spectra at the first two sensors are plotted in figure \ref{fig:fig06}$(a.i-a.ii)$, and the downstream evolution of the wall-pressure intensities are shown in panel $(b)$.  In these figures, light to dark blue curves corresponds to the DNS predictions from the initial linear estimate $\boldsymbol{\widetilde{c}}_0$, and two of the EnVar iterations, namely $\boldsymbol{\widetilde{c}}_1$ and the optimal $\boldsymbol{\widetilde{c}}_4$.  The DNS prediction using the initial linear estimate $\boldsymbol{\widetilde{c}}_0$ accurately reproduces the spectra at the first sensor within the frequency range $f \in \left[50, 350\right]$\,kHz\textemdash these are the inflow frequencies.  The DNS also shows the amplification of higher harmonics, $f \in \left[350, 600\right]$\,kHz, which indicates that the first sensor is already within the nonlinear regime.
This assertion was verified by evaluating the bicoherence of the wall-pressure data, both from the experimental and simulation data.
At sensor $s_2$, the spectra are relatively poorly predicted using $\boldsymbol{\widetilde{c}}_0$, which underscores the need for the nonlinear assimilation.  Specifically, the spectral peak near $f \in \left[250, 300\right]$\,kHz is over-predicted, and similarly is the intensity in figure \ref{fig:fig06}$(b)$. 
The impact of the EnVar optimization on the spectra at sensors $s_1$ and $s_2$ is most visible in the low-frequency range $f 
< 150\,\textrm{kHz}$, and appears to reduce the agreement with the measurements. The largest change is, however, near the spectral peak because the figure is in logarithmic scale, and appreciably improves the agreement with the experimental data.  
Recall that $\boldsymbol{\widetilde{c}}_4$ has less energy in the planar waves (figure \ref{fig:fig05}$b$) which explains the reduction in the spectra at $s_2$, and also the reduction of the intensity at that sensor as shown in figure \ref{fig:fig06}$(b)$.  To achieve this reduction, without compromising the accuracy of prediction at the first sensor, the disturbance intensity at the inflow was increased by including more energy in stable oblique waves (see figures \ref{fig:fig05}$b-c$). 

The dashed line in figure \ref{fig:fig06}$(b)$ is the prediction from the final assimilated state $\boldsymbol{\widetilde{c}}_4$, computed on grid G2 and compared to the data from sensors $s_3$ through $s_7$ which were not included in the assimilation.  Despite relatively accurate predictions at the first two sensors, the downstream prediction accuracy is very poor, even as early as at sensor $s_3$.  The discrepancies between measurements and simulations, especially near the recirculation bubble, demonstrate that the upstream two sensors are insufficient, and that further adjustments to $\boldsymbol{\widetilde{c}}_4$ that take into account downstream measurements should be considered.  

\subsection{Assimilation of the seven sensors data}
\label{sec:results_fine}%

\begin{figure}
    \centering
    \includegraphics[width=\textwidth]{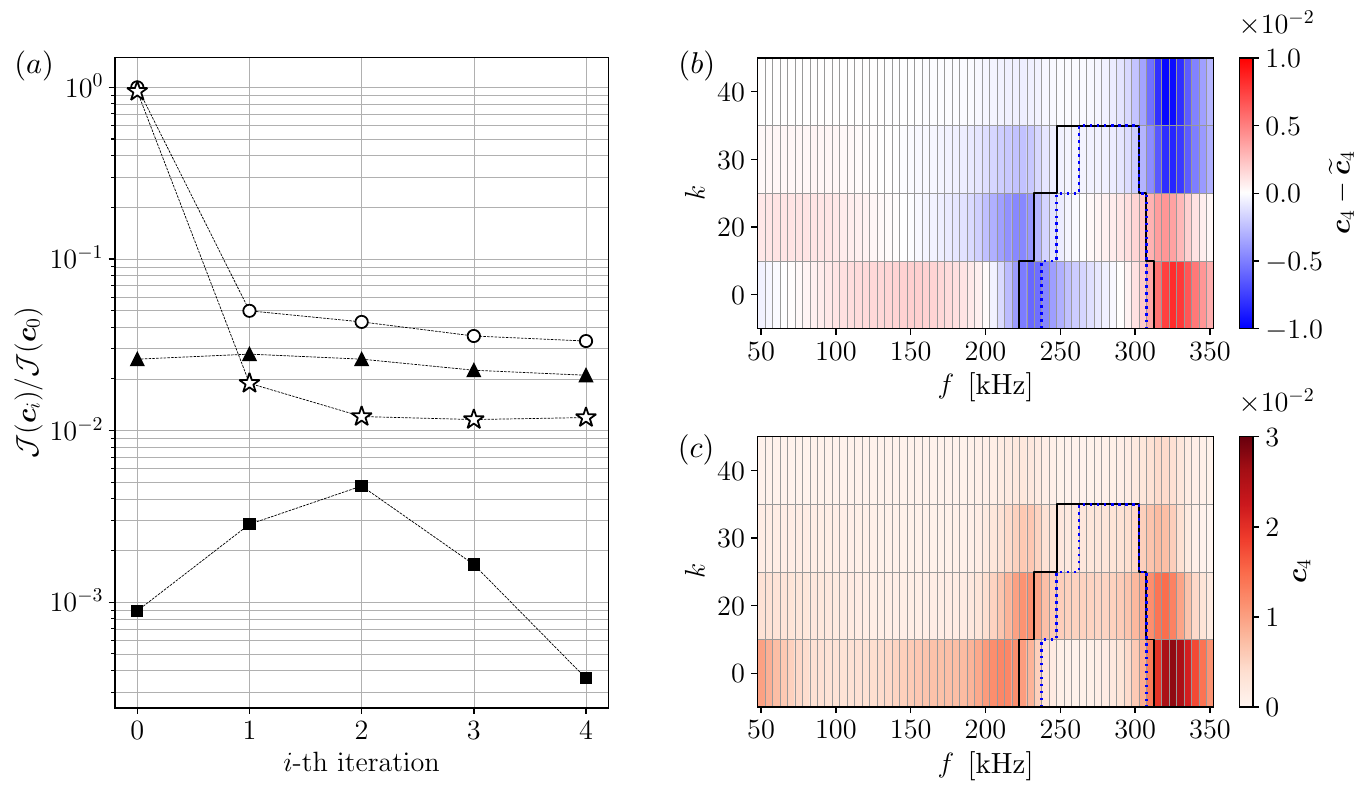} 
    \caption{EnVar assimilation of all seven sensors, on grid G2. (\textit{a}) Terms in the cost function normalized by the initial total cost $\mathcal{J}(\boldsymbol{{c}}_0)$:  
    ({\protect\colorcircle{1}{1}{1}{0}{0}{0}}) $\mathcal{J}$; 
    ({\protect\colortriangle{0}{0}{0}{0}{0}{0}}) $\mathcal{J}_{\scriptscriptstyle S}$; 
    ({\protect\colorstar{1}{1}{1}{0}{0}{0}}) $\mathcal{J}_{\scriptscriptstyle I}$; 
    ({\protect\colorsquare{0}{0}{0}{0}{0}{0}}) $\mathcal{J}_{\scriptscriptstyle P}$. 
    (\textit{b}) Difference between the spectra of the final assimilated control vector and its initial estimate ($\boldsymbol{c}_0 = \widetilde{\boldsymbol{c}}_{4}$). (\textit{c}) Spectra of the final assimilated control vector. Lines in ($b$) and ($c$) mark linearly unstable modes at ({\protect\dottedLine{0}{0}{1}}) inflow and ({\protect\solidLine{0}{0}{0}}) according to the $N$-factor on the cone.
}
    \label{fig:fig07}
    \vspace*{-4pt}
\end{figure}

\begin{figure}
    \centering
    \includegraphics[width=0.85\textwidth]{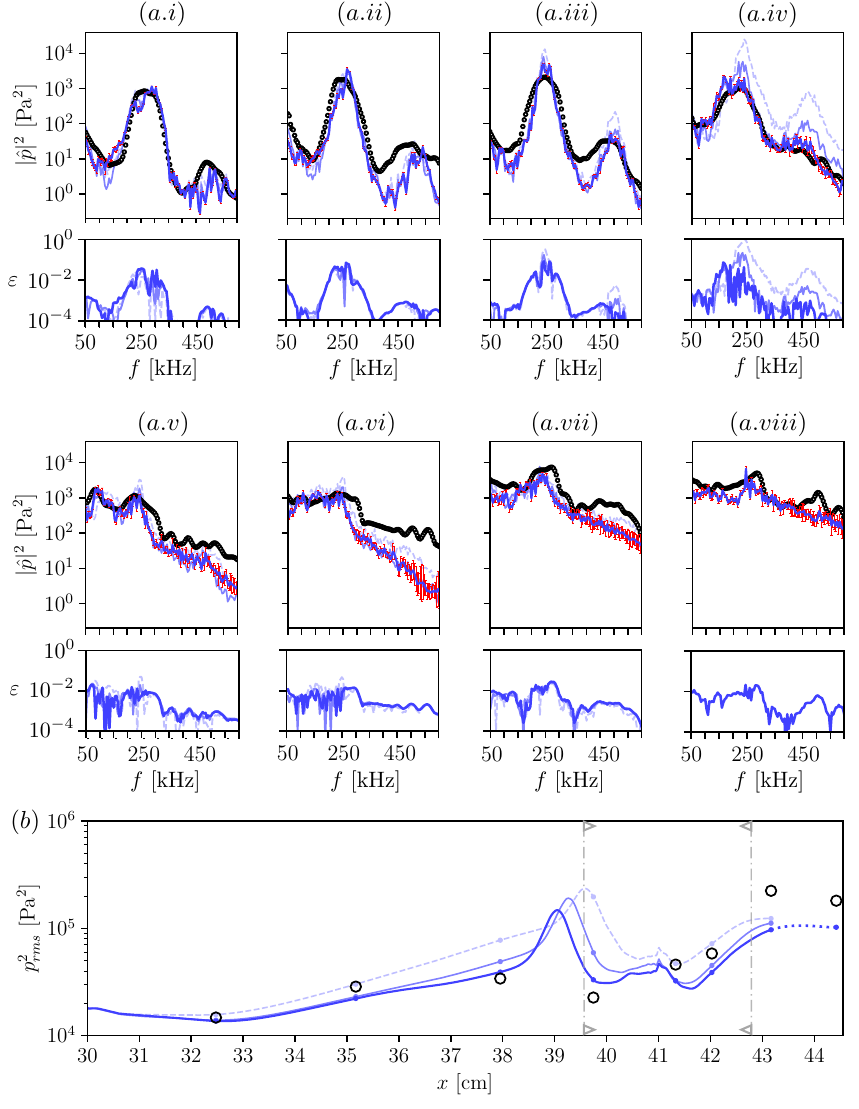}
    \caption{Wall-pressure spectra and intensity when assimilating all seven sensors. 
    (\textit{a.i}-\textit{a.viii}) Top panels are the spectra at sensors $s_1$-$s_8$; Bottom panels show the normalized errors $\varepsilon$. Sensor $s_{8}$ is not used in the assimilation. The error $\varepsilon(f)$ is defined as the absolute difference between the assimilated and experimental spectra, normalized by the intensity of the experimental data $\sum_f |\hat{p}|^2$. Red bars ({\protect\solidLine{1}{0}{0}}) are the variance $\sigma_{i}$ in the spectra for $5\%$ uncertainty in the assimilated flow $\boldsymbol{c}_{4}$. (\textit{b}) Intensity as a function of $x$. 
    Dashed line ({\protect\dashedLine{0}{0}{1}}) corresponds to $\boldsymbol{c}_{0} = \boldsymbol{\widetilde{c}}_{4}$ and is reproduced from figure \ref{fig:fig06}.  
    Black circles ({\protect\colorcircle{1}{1}{1}{0}{0}{0}}) indicate experimental measurements, solid lines ({\protect\solidLine{0}{0}{1}}) denote simulation results, and blue circles ({\protect\colorcircle{0}{0}{1}{0}{0}{1}}) mark intensities at sensor locations.  
    Light-to-dark blue represents EnVar iterations zero, one, and four.
    For the final estimate, the dotted extension ({\protect\dottedLine{0}{0}{1}}) between $s_{7}$ and $s_{8}$ signifies that the latter sensor was not part of the assimilation. 
    The vertical lines mark the locations of separation ({\protect\colortriangleleft{1}{1}{1}{0.75}{0.75}{0.75}}) and reattachment ({\protect\colortriangleright{1}{1}{1}{0.75}{0.75}{0.75}}) in the experiment.}
    \label{fig:fig08}
    \vspace*{-4pt}
\end{figure}%

The optimal control vector $\boldsymbol{\widetilde{c}}_{4}$ from the assimilation of the first two sensor data is adopted as an initial estimate, $\boldsymbol{c}_0= \boldsymbol{\widetilde{c}}_{4}$, for a new assimilation task, where the measurements from all seven sensors are considered.  All the simulations in this case are performed on grid G2.  
The convergence of the cost function is shown in figure~\ref{fig:fig07} for four EnVar iterations, during which the normalized cost reduces by approximately one and a half orders of magnitude. This reduction is dominated by the term containing the intensities, hinting that the optimization has acted to adjust the large difference in intensity within the recirculation bubble that was reported in figure \ref{fig:fig06}$(b)$.

The difference between the optimal control vector $\boldsymbol{c}_4$ and $\boldsymbol{c}_0 \left(=\boldsymbol{\widetilde{c}}_{4}\right)$ is shown in figure \ref{fig:fig07}$(b)$.  
The amplitudes of the unstable planar waves are reduced, while the three-dimensional unstable mode with azimuthal wavenumber $k=20$ shows a slight increase. These modification should be such that they do not compromise the agreement at sensors $s_1$ and $s_2$, yet improve the accuracy of reproducing the measurements  downstream of the second sensor where the original assimilation led to over-predictions.
In effect, the wall-pressure intensity at sensor three should be reduced, yet the intensity at the first two sensors should not change.  The tension between these two requirements can explain the increase in the amplitudes of some of the stable waves, including three-dimensional ones, that ultimately decay with distance but may be essential to match the early sensors. 
Also note that the increase in the amplitude of the unstable modes with $k=20$ in frequencies $f > 250\,\mathrm{kHz}$, which become stable downstream. 
The final control vector, $\boldsymbol{c}_4$, is shown in figure~\ref{fig:fig07}c, and the associated flow will be the focus of subsequent analysis.

Figure~\ref{fig:fig08} illustrates the impact of the assimilation of all seven sensors data on the accuracy of predicting the wall-pressure measurements. A comparison of the spectra from $\boldsymbol{c}_0 (=\boldsymbol{\widetilde{c}}_{4})$ and $\boldsymbol{c}_4$ shows clear changes starting at sensor $s_3$ and the most appreciable differences at sensor $s_4$.  
The figure also shows the frequency-dependent error $\varepsilon(f)$, defined as the difference between the assimilated and experimental spectra and normalized by the intensity $\sum_f |\hat{p}|^2$ of the experimental data.
We focus on $s_4$: The outcome of the first assimilation $\boldsymbol{c}_0 (=\boldsymbol{\widetilde{c}}_{4})$ yields a significant over-prediction at the two peaks in the range $f > 250\,\mathrm{kHz}$, which includes inflow-forcing frequencies and also nonlinearly generated harmonics. The further optimized control vector $\boldsymbol{c}_4$ reduces the spectra appreciably across this range without compromising the accuracy of lower ones, or the accuracy of predicting the spectra at the earlier sensors.  The wall-pressure intensity is shown in figure \ref{fig:fig08}$(b)$.   A minor mismatch is present in the data at sensor $s_2$, which is followed by a significant improvement in reproducing the intensity at sensors $s_3$-$s_4$.  
Between these two sensors, a large peak in intensity is predicted, which is not captured by the limited experimental measurements and which will be examined in \S\ref{sec:results_shock}.

As an independent assessment of the fidelity of the estimated flow, figure~\ref{fig:fig08} also includes a comparison to the wall-pressure spectra and intensity at sensor $s_{8}$, which is the first sensor downstream of the seven probes that were considered in the data assimilation.  Overall, the predicted spectra at $s_8$ approach the experimental measurements.
Compared to sensor $s_7$, the spectra at $s_8$ are similar in accuracy as quantified by $\varepsilon$, which indicates that the estimated upstream flow provides a realistic downstream state beyond the assimilated region. This behavior is consistent with the flow approaching a turbulent regime downstream, where different transition routes lead to statistically similar flows.

Whether the assimilation adopts the data from the first two sensors only or from all the probes, discrepancies persist in the prediction of the wall-pressure spectra and intensity at sensors six and seven.  The discrepancies in the spectra are primarily at high frequencies.  In order to ascertain whether this mismatch is due to sensitivity at the optimal state $\boldsymbol{c}_4$, we performed uncertainty quantification.  Assuming $5\%$ uncertainty in the posterior distribution of the optimal inflow vector $\boldsymbol{c}_4$, we can use the propagation of the final ensemble to compute the uncertainty in the spectra as 
$\boldsymbol{\sigma}^2 = \textrm{diag}\left(\boldsymbol{\Sigma}_{{\scriptscriptstyle H_{4}}}\right)  (0.05 |\!|\boldsymbol{c}_{4}|\!|_{\scriptscriptstyle \infty})^2/|\!| \boldsymbol{\Sigma}_{\boldsymbol{c}_4}|\!|_2 $, where $\textrm{diag}\left(\boldsymbol{\Sigma}_{{\scriptscriptstyle H_{4}}}\right)$ are the diagonal elements of the observation covariance matrix $\boldsymbol{\Sigma}_{{\scriptscriptstyle H_{4}}} \! = \! \mathsfbi{HH}^{\top}\! /(N_{ens}-1)$.  
These uncertainty bands are marked in red in figure \ref{fig:fig08}$a$.  
While uncertainties bands expand with  downstream disturbance, in particular at the final two sensors, they do not account for the deviation from the experimental measurements, and an explanation will be provided in \S\ref{sec:results_mismatch}.

\begin{figure}
    \centering
    \includegraphics[width=\textwidth]{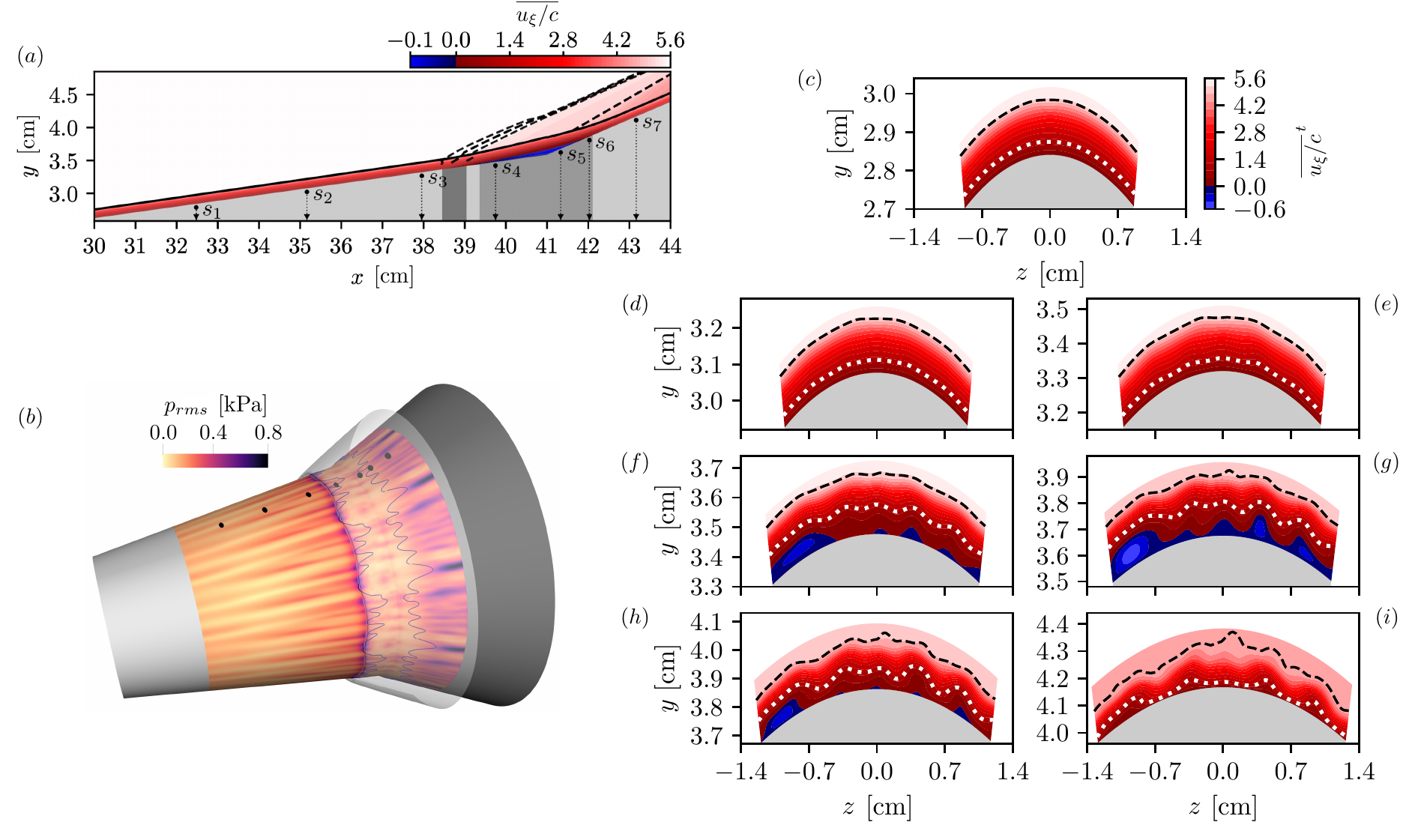}
    \caption{
    Mean assimilated flow state, $\boldsymbol{q} = \mathcal{N}(\boldsymbol{c}_{4})$. (\textit{a}) Contours of time and azimuthally averaged streamwise Mach number. Black solid line (\protect\solidLine{0}{0}{0}) marks the boundary-layer edge $\overline{\delta}_{\scriptscriptstyle 99}$. Black dashed lines (\protect\dashedLine{0}{0}{0}) identify the separation and reattachment shocks using the conditions, $\Upsilon(x,y) = \{1, 0.5\}$ (equation~(\ref{eq:shock_identification}); dark gray area $x=[38.5,\,39.0]\,\mathrm{cm}$ is the extent of the shock foot; gray shaded area $x=[39.4,\,42.1]\,\mathrm{cm}$ is the extent of separation (table~\ref{tab:separation}).
    (\textit{b}) Contours of the root-mean-squared wall pressure, computed with respect to time only.  Solid lines (\protect\solidLine{0}{0}{0}) mark separation and reattachment, $\Gamma(x,\vartheta) = 0.5$. The white isosurface shows the mean separation shock, generated by revolving the curve $\Upsilon(x,y)=1$ around the x-axis.
    (\textit{c}-\textit{i}) Contours of time-averaged streamwise Mach number, on the vertical planes above sensors $s_{1}$ through $s_{7}$. The boundary-layer edge is marked by a dashed line (\protect\dashedLine{0}{0}{0}), and the sonic line is shown with a white dotted line (\protect\dottedLine{0}{0}{0}).
    }
    \label{fig:fig09}
    \vspace*{-4pt}
\end{figure}

Figure~\ref{fig:fig09} shows the assimilated mean flow $\boldsymbol{\overline{q}}$ from the final solution $\boldsymbol{c}_4$, where $\boldsymbol{q}= \mathcal{N}(\boldsymbol{c}_4)$. 
The contours in panel ($a$) are the time-and-azimuthally averaged streamwise velocity normalized by the speed of sound.  
On the figure, dashed lines mark the separation shock which is identified in a manner similar to \cite{lovely1999_shock} using the criterion,
\begin{equation}
   \Upsilon(x,y) = \overline{\left| \boldsymbol{u} \cdot \nabla p \right| / (c|\!| \nabla p|\!|_2)}  = 1,
\label{eq:shock_identification}
\end{equation}
where $\nabla p$ is the pressure gradient. 
Starting in the free stream and approaching the edge of the boundary layer, the separation-shock thickness increases due to the gradual turning of the flow and the unsteadiness of the shock-boundary layer interaction.  Below the boundary-layer edge, $\delta_{\scriptscriptstyle 99}$, the shock foot further widens down to the location where the flow becomes (sub)sonic. Beneath the shock foot (dark shaded region in figure \ref{fig:fig09}($a$)), the wall-pressure fluctuation intensity undergoes a significant amplification as shown in the earlier figure~\ref{fig:fig08}($b$). This peak intensity is not captured by the experimental placement of the sensors, and will be attributed to the amplification of disturbances as they traverse the region beneath the shock foot.

The locations of separation and reattachment are evaluated using an intermittency function based on the sign of the tangential wall shear-stress $\tau_{w}$.  Specifically, we identify the iso-level, 
\begin{equation}
    \Gamma (x, \vartheta) = \overline{\gamma(x,\vartheta,t)}^{\, t} = 0.5, \quad
    \textrm{where}\quad \gamma(x,\vartheta,t) = \left\{
\begin{matrix}
0, \ \mathrm{if} \ \tau_{w}(x,\vartheta,t) \geq 0\\
1, \ \mathrm{if} \ \tau_{w}(x,\vartheta,t) < 0 \\
\end{matrix}
\right. .
\label{eq:separation}
\end{equation}
The locations of separation and reattachment identified by this threshold are then azimuthally averaged, 
$\overline{x_s}$, $\overline{x_r}$, and are marked by the light shaded region in figure \ref{fig:fig09}$(a)$. 
The precise coordinates are reported in table~\ref{tab:separation} alongside the experimental values.  The latter were determined from schlieren images by \citet{butler2022_coneflare} who identified changes in the slope of the direction along which disturbances appear to propagate. 
The agreement between the simulations and the experiments is evidence of the success of the assimilation procedure, which did not incorporate any information regarding this observable in the cost function. Separation and reattachment location from the simulations fall within the $99\%$ confidence interval of the reported experimental data. Note, however, that azimuthally averaging the separation and reattachment locations conceals their azimuthal variations which is shown in figure~\ref{fig:fig09}(\textit{b}). The figure also shows the streaky patterns of the root-mean-squared wall pressure, which is not captured by the available experimental measurements.

\begin{table}
    \centering
    \begin{tabular}{lll}
        &  Separation & Reattachment  \\[0.4em] 
        
        Laminar, $\boldsymbol{q}_{B}$  & $38.5\,\mathrm{cm} \hspace*{23pt}$ & $42.9\,\mathrm{cm}$ \hspace*{20pt} \\
        Assimilated, $\boldsymbol{q}=\mathcal{N}(\boldsymbol{c}_4)$  & $39.4\,\pm\,0.4\,\mathrm{cm} \hspace*{23pt}$ & $42.1\,\pm\,0.4\,\mathrm{cm}$ \hspace*{20pt} \\
        Experiments   & $39.6\,\pm\,0.3\,\mathrm{cm}$ & $42.8\,\pm\,0.3\,\mathrm{cm}$ \hspace*{20pt} \\
    \end{tabular}
    \caption{Separation and reattachment locations in the experiment and the simulations. 
    Simulations results are the azimuthal averages of the values from equation~(\ref{eq:separation}), $\overline{x_s}$ and $\overline{x_r}$, plus/minus one standard deviation. Experimental results based on the change in direction of disturbance propagation \citep[see][]{butler2021_phdthesis}.}
    \label{tab:separation}
\end{table}

The time-averaged local Mach number based on the streamwise velocity, $\overline{ u_{\xi}/ c }^{\, t}$, is plotted in the vertical planes above the sensor locations in figures~\ref{fig:fig09}($c$-$i$). The averaged flow is azimuthally homogeneous at the first three sensor locations, and shows clear streaky patterns at the fourth sensor, which is within the recirculation bubble. This pattern is consistent with earlier studies \citep{dwivedi2019_reattachment, paredes2022_cone} that examined the amplification of energetic three-dimensional structures when boundary-layer disturbances interact with recirculation regions.

The intensification of the root-mean-square wall pressure beneath the separation shock is followed by a fast decay within a narrow region between sensors $s_3$ and $s_4$ (see figure \ref{fig:fig08}($b$) and \ref{fig:fig09}($a$,$b$)). Although these sensors do not capture the peak intensity, they are instrumental in enabling EnVar optimization to correctly localize this phenomenon within the simulations. This finding suggests that, when a recirculation bubble is present, experimental data collection would benefit from a greater number or optimized placement of sensors near the separation point.

In summary, the results of the EnVar assimilation align satisfactorily with the experimental data, although some discrepancies remain visible. The simulations nonetheless unveil new features and provide full details of the flow that we will examine further.
First, the root-mean-square wall pressure intensifies sharply beneath the separation shock, which was not captured by the experiments. Second, as expected, the separation bubble alters the disturbance spectra.  Finally, the discrepancy between the simulation and experimental data at the last two sensors points to areas requiring further investigation. These three aspects will be explored in the following sections.

\subsection{Disturbance behavior across the shock--boundary layer interaction}
\label{sec:results_shock}%

\begin{figure}
    \centering
    \includegraphics[width=\textwidth]{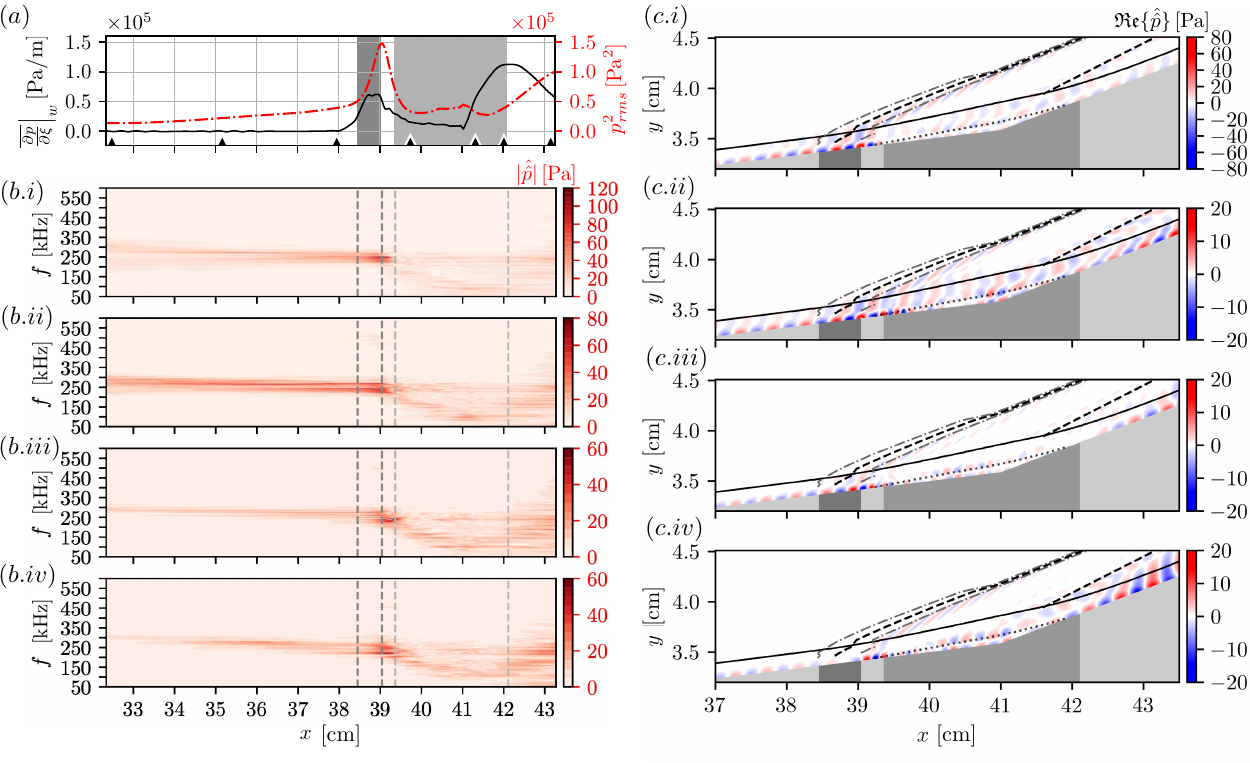} 
    \caption{Pressure data from the assimilated flow, $\boldsymbol{q} = \mathcal{N}(\boldsymbol{c}_{4})$. (\textit{a}) Time and azimuthally averaged ({\protect\solidLine{0}{0}{0}}) streamwise gradient of the wall pressure and ({\protect\dashedDottedLine{1}{0}{0}}) mean-squared wall-pressure fluctuations; dark gray area $x=[38.5,\,39.0]\,\mathrm{cm}$ is the extent of the shock foot; gray shaded area $x=[39.4,\,42.1]\,\mathrm{cm}$ is the extent of separation. 
    (\textit{b}) Amplitudes of the $(f, k)$ Fourier coefficients of the wall pressure; dashed lines mark the boundaries of the shock foot ({\protect\dashedLine{0.4}{0.4}{0.4}}) and separation ({\protect\dashedLine{0.7}{0.7}{0.7}}).  
    (\textit{c}) Pressure Fourier modes at $f=250\,\mathrm{kHz}$; (\protect\solidLine{0}{0}{0})  $\overline{\delta}_{\scriptscriptstyle 99}$; (\protect\dashedLine{0}{0}{0}) $\Upsilon(x,y) = 1$; (\protect\dashedLine{0.4}{0.4}{0.4}) $\Upsilon(x,y) = 0.5$; (\protect\dottedLine{0}{0}{0}) $u_{\xi} = 0$; dark and light gray areas as in (\textit{a}). (\textit{i}-\textit{iv}) $k=\{0, 20, 30, 40\}$.}
    \label{fig:fig10}
    \vspace*{-4pt}
\end{figure}

The root-mean-squared wall pressure, $p_{rms}^2$, is plotted in figure~\ref{fig:fig10}($a$) as a function of downstream distance.  The figure also shows the average streamwise pressure gradient, $\overline{\partial p / \partial \xi}$.  It is evident that the intensification of $p_{rms}$ is correlated with the adverse pressure gradient beneath the shock foot, and peaks at the end of this region.

A more detailed view is provided in figure \ref{fig:fig10}($b$), where contours of the wall-pressure frequency spectra are plotted as a function of the downstream distance, $|\widehat{p}(f,x)|$. The four panels correspond to azimuthal wavenumbers $k=\{0, 20, 30, 40\}$.  The figure shows that the adverse-pressure-gradient region amplifies all the oncoming disturbances, without a distinct preference for specific frequencies.  As such, the contours are dominated by the disturbances that are most amplified from upstream, and which are further amplified under the shock foot.  
These are the unstable Mack modes with frequency $f=250\,\mathrm{kHz}$, which dominate the spectra upstream of the separation shock, and remain dominant beneath the shock foot where they undergo further amplification.  The associated mode shapes are shown in figure \ref{fig:fig10}($c$), where contours of $\mathfrak{Re}\{\widehat{\widehat{p}}\}$ are plotted at $f=250\,\mathrm{kHz}$ and $k=\{0, 20, 30, 40\}$.  
The pressure disturbance is largely confined within the boundary layer, primarily below the relative sonic line \citep{fedorov2011_transition}. 
The shock amplifies this near-wall portion of the disturbance energy appreciably.  
The portions of the pressure disturbances that are located above the relative sonic line, though lower in amplitude, are amplified to a lesser degree as they traverse the separation shock.  In addition, these disturbances are deflected and radiate, or propagate, along the shock inducing fluctuations in the shock itself.

%
\begin{figure}
    \centering
    \includegraphics[width=.7\textwidth]{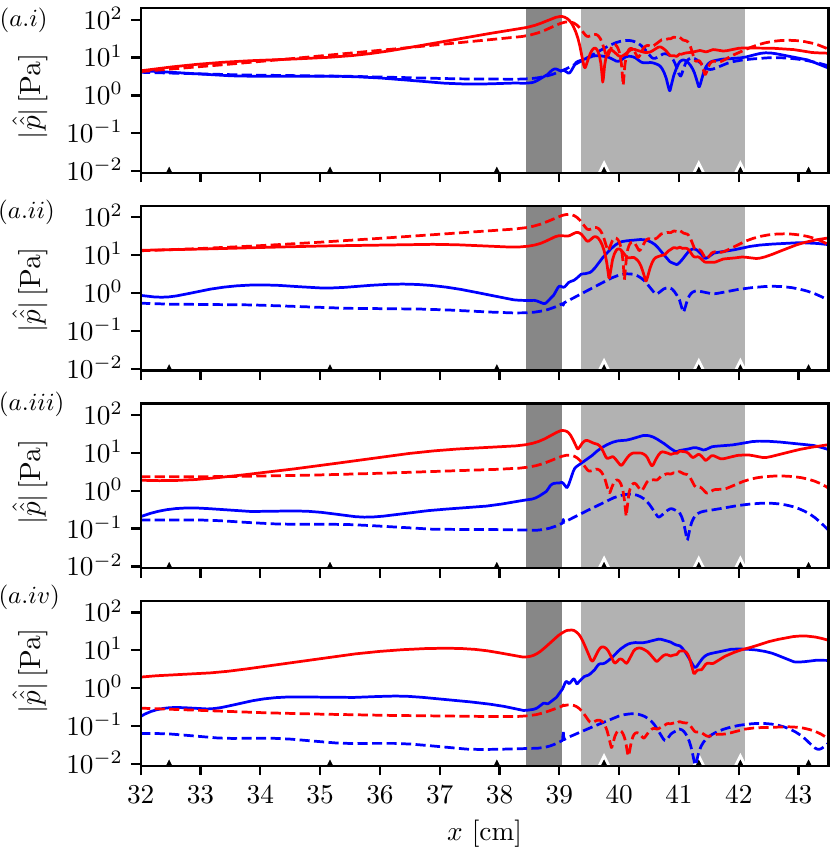} 
    \caption{Nonlinear and linear development of particular $(f, k)$ Fourier components of the wall pressure, for the assimilated inflow $\boldsymbol{c}_{4}$.   
    (Solid) Nonlinear Navier-Stokes solution $\boldsymbol{q}=\mathcal{N}(\boldsymbol{c}_{4})$ ({\protect\solidLine{1}{0}{0}}), ({\protect\solidLine{0}{0}{1}}); 
    (dashed) linearized Navier-Stokes solution $\boldsymbol{q}_{\scriptscriptstyle \mathcal{L}}^{\prime} \! = \! \mathcal{L}_{\overline{\boldsymbol{q}}}(\boldsymbol{c}_{4})$ ({\protect\dashedLine{1}{0}{0}}),({\protect\dashedLine{0}{0}{1}}).
    (Red) $f=250\,\mathrm{kHz}$; (blue) $f=150\,\mathrm{kHz}$;
    (\textit{a}.i-iv) $k = \{0, 20, 30, 40\}$.
    Dark gray area $x=[38.5,\,39.0]\,\mathrm{cm}$ is the extent of the shock foot; light gray area $x=[39.4,\,42.1]\,\mathrm{cm}$ is the extent of separation.
    }
    \label{fig:fig11}
    \vspace*{-4pt}
\end{figure}
\begin{figure}
    \centering
    \includegraphics[width=.65\textwidth]{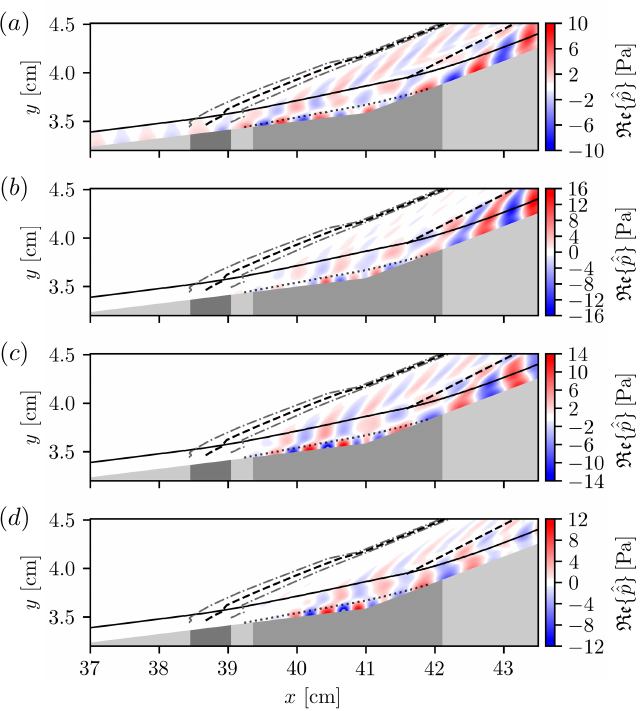}
    \caption{Fourier modes of the final assimilated field, at $f=150\,\mathrm{kHz}$ and ($a$-$d$) $k=\{0, 20, 30, 40\}$. Dark gray area $x=[38.5,\,39.0]\,\mathrm{cm}$ is the extent of the shock foot; light gray area $x=[39.4,\,42.1]\,\mathrm{cm}$ is the extent of separation. (\protect\solidLine{0}{0}{0})  $\overline{\delta}_{\scriptscriptstyle 99}$; (\protect\dashedLine{0}{0}{0}) $\Upsilon(x,y) = 1$; (\protect\dashedLine{0.4}{0.4}{0.4}) $\Upsilon(x,y) = 0.5$; (\protect\dottedLine{0}{0}{0}) $u_{\xi} = 0$.}
    \label{fig:fig12}
    \vspace*{-4pt}
\end{figure}
The amplification of the boundary-layer disturbances beneath the shock foot, in particular the dominant modes near $f \sim 250\,\mathrm{kHz}$, is followed by a fast decay within the separation region, primarily between the end of the shock foot and the onset of the recirculation bubble (figures~\ref{fig:fig10}($a$,$b$)). This decay does not affect all frequencies equally, however. Upon leaving the shock, the amplitudes of the dominant Mack modes decrease, while lower-frequency modes ($f < 150\,\mathrm{kHz}$) begin to amplify. Within the recirculation bubble, the Mack modes retain their weakened amplitudes. As for the lower-frequency waves that grow within the bubble, the three-dimensional ones ($k \! \neq \! 0$) experience greater amplification than their two-dimensional counterparts ($k \! = \! 0$). These trends are consistent with results from earlier linear stability analysis \citep[e.g.][]{paredes2022_cone}, which predicts low-frequency three-dimensional modes as unstable and Mack modes neutrally stable within the bubble.

In figure~\ref{fig:fig11}, we report the evolution of the wall pressure at particular frequency-wavenumber pairs, evaluated using spectral analysis of the assimilated flow, which is a solution of the nonlinear Navier-Stokes equations $\boldsymbol{q} \! = \! \mathcal{N}(\boldsymbol{c}_{4})$.  For comparison, we also plot the linear evolution of the same frequency-wavenumber pairs, computed using the linearized Navier-Stokes operator and the mean state, $\boldsymbol{q}_{\scriptscriptstyle \mathcal{L}}^{\prime} \! = \! \mathcal{L}_{\overline{\boldsymbol{q}}}(\boldsymbol{c}_{4})$.  
The originally unstable frequencies (red curves, $f\!=\!250\,\mathrm{kHz}$) show the fast decay upstream of separation, both in the nonlinear and linear computations. 
More importantly, consider the low-frequency and three-dimensional modes (blue, $f\!=\!150\,\mathrm{kHz}$ and $k \! \neq \! 0$).  For these disturbances, the nonlinear evolution shows a stronger amplification within the bubble compared to the linear counterpart.  
The comparison thus underscores the importance of studying the nonlinear assimilated field, where separation onset closely reproduces the experimental location.  

Contours of the pressure disturbances associated with the low-frequency modes ($f\!=\!150\,\mathrm{kHz}$) are shown in figure \ref{fig:fig12}, where the real components of the Fourier representation are plotted for $k=\{0, 20, 30, 40\}$.  
The highest amplitudes are mostly located within the recirculation bubble where the flow direction is reversed (below the dotted line).  Outside the bubble, within the forward boundary-layer flow, the amplitudes of the modes are lower and the phase is reversed. 
Downstream of reattachment, an appreciable change in the contours is again observed, both in terms of the disturbance profiles and their amplitudes.  The increase in amplitude is consistent with the change in the spectra of the low-frequency modes as we approach sensor $s_7$  (see figure \ref{fig:fig08}($a$)).


\subsection{Mismatch at the final sensor position}
\label{sec:results_mismatch}

We now revisit the spectra in figure~\ref{fig:fig08}($a$) and focus on the final two sensors $s_6$ and $s_7$, along the flare.  
These spectra capture an appreciable amplification in the high-frequency modes with $f\!\gtrsim\!250\,\mathrm{kHz}$.  However, discrepancies arise between the simulated and experimental spectra, especially in the frequency range $f\!>\!300\,\mathrm{kHz}$.  The rapid increase in the energy spectral density in this range, relative to the upstream sensors, is itself an important hint.  This amplification of the high frequencies will be shown to depend on the extent of the separation bubble, which is an important source of uncertainty.

Separation and reattachment in this flow are neither steady nor azimuthally invariant.  Unlike the experiments, where the sensor is at one azimuthal location and averaging is performed in time only, in the simulations we also averaged the observations from all azimuthal locations.  Since $s_6$ is within $1\,\textrm{mm}$ of reattachment in the assimilated field, our averaging samples points include both pre- and post-reattachment. Absent azimuthal probes, this was deemed to be the best approach to remove bias when analyzing the assimilated fields. Here we will focus on the temporal dynamics, since the unsteadiness of separation and reattachment influences both the single-point data from the experiments and also the simulations.

\begin{figure}
    \centering
    \includegraphics[width=.98\textwidth]{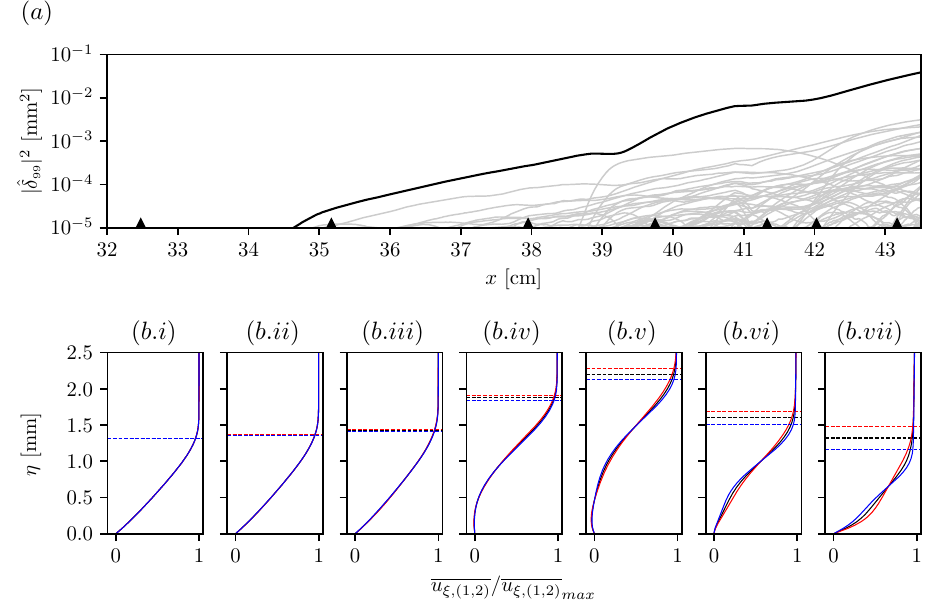} 
    \caption{Spectra of the boundary-layer thickness and mean streamwise-velocity profiles of the assimilated flow $\boldsymbol{q} = \mathcal{N}(\boldsymbol{c}_{4})$.
    (\textit{a}) Streamwise evolution of Fourier components $\hat{\delta}_{99}$ at $f=\{1,2,\dots,600\}\,\mathrm{kHz}$ in gray ({\protect\solidLine{0.65}{0.65}{0.65}}), with the dominant $f=5\,\mathrm{kHz}$ in black ({\protect\solidLine{0}{0}{0}}).
    ($b.i$–$b.vii$) Normalized profiles of the time and azimuthally averaged streamwise velocity ($\overline{u}_{\xi}$) at sensors $s_{1}$ to $s_{7}$ ({\protect\solidLine{0}{0}{0}}), ({\protect\solidLine{0}{0}{1}}), ({\protect\solidLine{1}{0}{0}}). Horizontal lines mark $\delta_{\scriptscriptstyle 99}$ ({\protect\dashedLine{0}{0}{0}}), ({\protect\dashedLine{0}{0}{1}}), ({\protect\dashedLine{1}{0}{0}}).
    Black: average over the full time horizon $5T=1\,\mathrm{ms}$ (where $T=1/(5\,\mathrm{kHz})=0.2\,\mathrm{ms}$);
    Red/Blue: conditional averages over $[t_0,t_0+T/2)$ and $[t_0+T/2,t_0+T)$, where $t_0$ in each panel is selected as the start of the positive phase of $\langle \delta_{\scriptscriptstyle 99} \rangle_{5 \, \mathrm{kHz}}$.}
    \label{fig:fig13_1kHz}
    \vspace*{-4pt}
\end{figure}

In order to examine the effect of the low-frequency unsteadiness that takes place in shock-boundary layer interactions, we consider the temporal variation of the boundary-layer thickness $\delta_{\scriptscriptstyle 99}(x,t)$ and of the streamwise velocity profile $\overline{u_{\xi}}^{\, \vartheta}$. Low-frequency oscillations in these quantities can affect the amplification of high-frequency disturbances. The frequency spectra of $\delta_{\scriptscriptstyle 99}$ are reported in figure~\ref{fig:fig13_1kHz}$(a)$, and show that $f=5\,\mathrm{kHz}$ is dominant, and grows by more than two orders of magnitudes from upstream of the second sensor to downstream of the last sensor. 
This low-frequency component does not originate from the inflow, as the spectral make-up of the inflow disturbance includes modes starting at $50\,\mathrm{kHz}$ with a $5\,\mathrm{kHz}$ resolution. We therefore attribute it to nonlinearity (recall that the simulation time series spans $1\,\mathrm{ms}$, and therefore the minimum frequency is $1\,\mathrm{kHz}$).  
This oscillation also manifests itself in the azimuthally averaged streamwise velocity, $\overline{u_{\xi}}^{\, \vartheta}$.  In figure \ref{fig:fig13_1kHz}$(b)$, we plot the conditional average of $\overline{u_{\xi}}^{\, \vartheta}$ over two half-periods of the $5\,\mathrm{kHz}$ cycle, specifically $\overline{u}_{\xi,1}$ over $[t_0,t_0+T/2)$ and $\overline{u}_{\xi,2}$ over $[t_0+T/2,t_0+T)$ where $T=1/(5\,\mathrm{kHz})=0.2\,\mathrm{ms}$, with $t_0$ chosen separately for each sensor as the start of the positive half-cycle of $\langle \delta_{\scriptscriptstyle 99} \rangle_{5\,\mathrm{kHz}}$ (i.e.\,$\langle \delta_{\scriptscriptstyle 99} \rangle_{5\,\mathrm{kHz}}$ is positive throughout the first sub-interval). The figure shows that the two streamwise-velocity profiles differ most noticeably at the last two sensors, where they are clearly thicker (red) and thinner (blue) than the unconditional mean (black), consistent with the low-frequency oscillation in $\delta_{\scriptscriptstyle 99}$.  

\begin{figure}
    \centering
    \includegraphics[width=.88\textwidth]{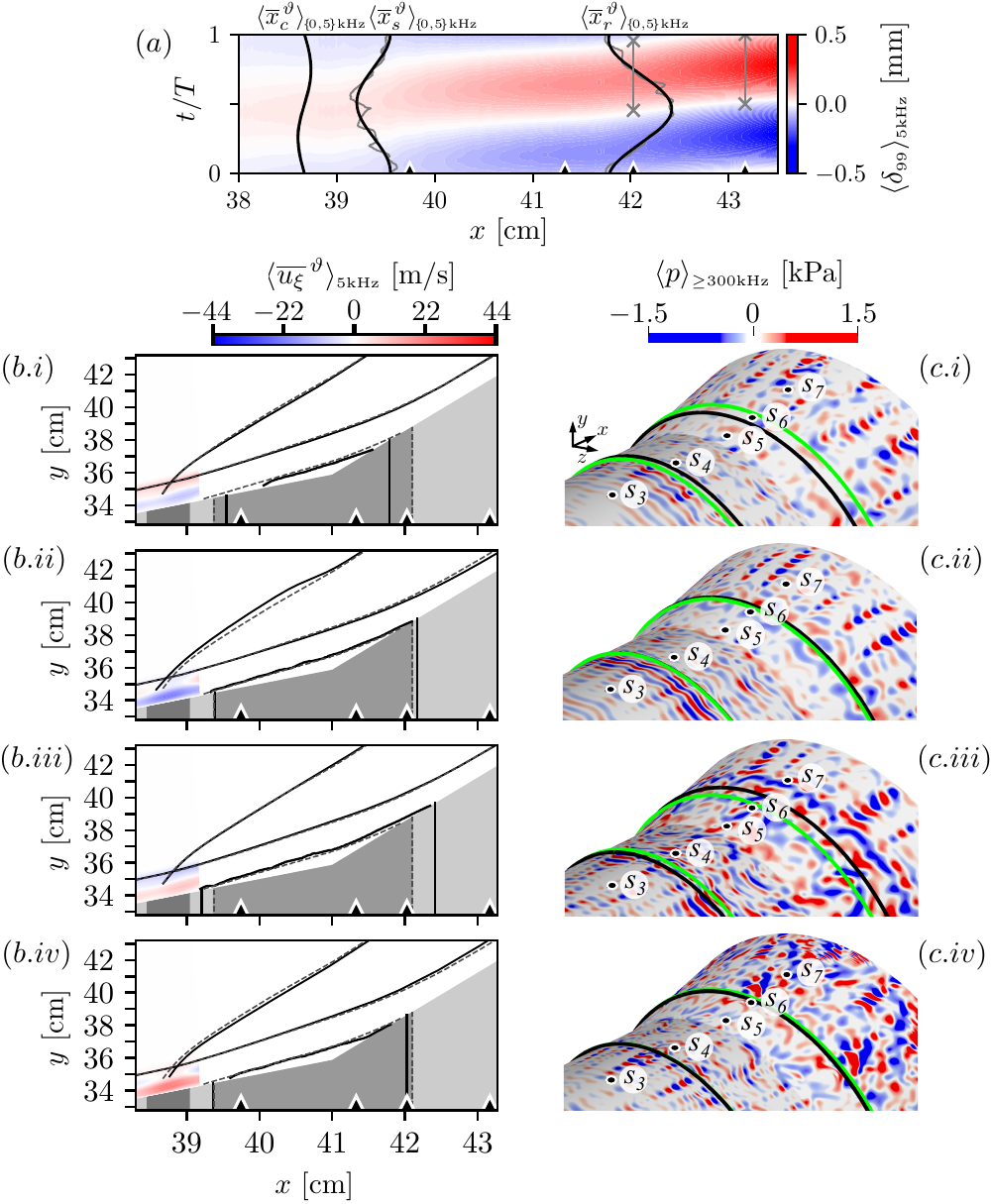} 
    \caption{Low-frequency unsteadiness in the assimilated state, $\boldsymbol{q} = \mathcal{N}(\boldsymbol{c}_{4})$. 
    (\textit{a}) $5\,\mathrm{kHz}$-filtered boundary-layer thickness, $\langle\delta_{99}\rangle_{5\,\mathrm{kHz}}$.  ({\protect\solidLine{0.65}{0.65}{0.65}}) Azimuthal averages $\overline{\bullet}^{\, \vartheta}$ and ({\protect\solidLine{0}{0}{0}}) time average plus the $5\,\mathrm{kHz}$ filtered $\langle \overline{\bullet}^{\,\vartheta}\rangle_{\{0,5\}\,\mathrm{kHz}}$ positions of: compression shock $x_{c}$; separation onset $x_{s}$; and reattachment $x_{r}$. 
    Crosses ({\protect\colorcross{0.65}{0.65}{0.65}}) mark the interval $[t_0, t_0 + T/2)$ used for conditional averaging of $\overline{u}_{\xi,1}$ in figure~\ref{fig:fig13_1kHz}($b$). 
    (\textit{b}) Snapshots during $5\,\mathrm{kHz}$ flow oscillation. Contours are velocity disturbances.  Lines are the corner shock, $\delta_{\scriptscriptstyle 99}$, and separation bubble.
    Solid ({\protect\solidLine{0}{0}{0}}): Azimuthal and time-averaged curves. Dashed ({\protect\dashedLine{0}{0}{0}}): Azimuthal averages at (\textit{i--iv}) $t/T = \{0,\,0.25,\,0.5,\,0.75\}$ during $5\,\mathrm{kHz}$ oscillation, 
    with $t=0$ chosen such that the shock is at the time-averaged position. 
    Black triangles ({\protect\colortriangle{0}{0}{0}{0}{0}{0}}) are sensors $s_4$ to $s_7$.
    Dark gray area $x=[38.5,\,39.0]\,\mathrm{cm}$ is the extent of the shock foot; light gray area $x=[39.4,\,42.1]\,\mathrm{cm}$ is the extent of separation. 
    (\textit{c}) High-pass-filtered wall pressure, $f\ge300\,\mathrm{kHz}$. 
    Green lines ({\protect\solidLine{0}{1}{0}}) are time-averaged $\overline{x}_s$ and $\overline{x}_r$; black lines ({\protect\solidLine{0}{0}{0}}) are instantaneous $\langle \overline{x}_s^{\,\vartheta} \rangle_{\{0,5\}\,\mathrm{kHz}}$ and $\langle \overline{x}_r^{\,\vartheta} \rangle_{\{0,5\}\,\mathrm{kHz}}$.
    Black circles ({\protect\colorcircle{0}{0}{0}{0}{0}{0}}) are sensors $s_{3}$ to $s_{7}$. (\textit{i--iv}) $t/T = \{0,\,0.25,\,0.5,\,0.75\}$.}
    \label{fig:fig14_1kHz}
    \vspace*{-4pt}
\end{figure}

Another representation of the unsteadiness in the boundary-layer thickness is shown in figure \ref{fig:fig14_1kHz}$(a)$. The contours are $\delta_{\scriptscriptstyle 99}$ filtered at the $5\,\mathrm{kHz}$ rate, plotted as functions of time and streamwise distance.  The crosses at sensors $s_6$ and $s_7$ identify the boundaries of the sub-intervals used in the conditional averaging, $t_0$ and $t_0+T/2$.  Superimposed on the contours are lines that correspond to the time-dependent positions of the corner shock $\overline{x}_c^{\, \vartheta}$, separation onset $\overline{x}_s^{\, \vartheta}$, and reattachment $\overline{x}_r^{\, \vartheta}$.  For each quantity, the azimuthal average ($\overline{\bullet}^{\, \vartheta}$) is plotted in gray. Also shown, in black, are the time averages plus the $5\,\mathrm{kHz}$ filtered signals, $\langle\overline{\bullet}^{\, \vartheta}\rangle_{\{0,5\}\, \mathrm{kHz}}$.  The filtered curves largely capture the observed unsteadiness of all three quantities, $\{\overline{x}_c^{\, \vartheta}, \overline{x}_s^{\, \vartheta}, \overline{x}_r^{\, \vartheta} \}$.  In fact, for the corner shock location $\overline{x}_c^{\, \vartheta}$ the oscillation of the filtered and unfiltered signals are nearly identical, making the unfiltered trace barely distinguishable in the figure.
For $\langle \overline{x}_s^{\, \vartheta} \rangle_{\{0,5\}\, \mathrm{kHz}}$, the upstream-most location of separation is in phase with the largest boundary-layer thickness (red contours), and the downstream-most $\langle \overline{x}_s^{\, \vartheta} \rangle_{\{0,5\}\, \mathrm{kHz}}$ occurs when the boundary layer is thinnest.  Reattachment $\langle \overline{x}_r^{\, \vartheta} \rangle_{\{0,5\}\, \mathrm{kHz}}$ is nearly exactly out of phase with separation $\langle \overline{x}_s^{\, \vartheta} \rangle_{\{0,5\}\, \mathrm{kHz}}$, thus leading to the longest bubble extent at early separation and the shortest extent when separation is delayed. 

The side views in panels $(b.i-b.iv)$ show four phases within the low-frequency cycle. A negative streamwise-velocity fluctuation appears beneath the shock at $t=0$ (figure~\ref{fig:fig14_1kHz}\textit{b.i}) and intensifies at $t=0.25\,T$ (figure~\ref{fig:fig14_1kHz}\textit{b.ii}), moving the foot of the compression shock ($\langle \overline{x}_{c}^{\, \vartheta} \rangle_{\{0,5\}\, \mathrm{kHz}}$) from its neutral position to its maximum retraction upstream. At $t=0.5\,T$, the shock foot ($\langle \overline{x}_{c}^{\, \vartheta} \rangle_{\{0,5\} \mathrm{kHz}}$) has returned to its neutral position, and the streamwise-velocity fluctuation is positive (figure~\ref{fig:fig14_1kHz}\textit{b.iii}). At $t=0.75\,T$, the streamwise-velocity fluctuation intensifies further in the positive direction, and the shock foot reaches its maximum downstream displacement (figure~\ref{fig:fig14_1kHz}\textit{b.iv}).  The figure also clearly captures the time delay in the movement of the separation point, which is shifted by a quarter period.  

The corresponding changes in the wall-pressure fluctuations above $300\,\mathrm{kHz}$ are shown in figure~\ref{fig:fig14_1kHz}($c$), where we plot the spectrally high-pass filtered pressure signal at the same four phases as in panel ($b$).  While this interpretation does not take into account the travel time of the perturbations, it is justified by a separation of timescales: 
The residence time within the separated region ($\overline{x}_{s} < x < \overline{x}_{r}$, or $39.4\,\mathrm{cm} < x < 42.1\,\mathrm{cm}$)
of disturbances with frequencies $50\,\mathrm{kHz}\leq f \leq 600\,\mathrm{kHz}$ is less than $0.038\,\mathrm{ms}$. This duration is approximately a factor of five shorter than the period of the dominant $5\,\mathrm{kHz}$ oscillation, $T = 0.2\,\mathrm{ms}$, which dominates the changes in the boundary-layer thickness and the bubble expansion–contraction cycle.
The lowest and highest pressure oscillations in this figure correspond to the retracting and expanding phases of the separation region, respectively.

\begin{figure}
    \centering
    \includegraphics[width=0.98\textwidth]{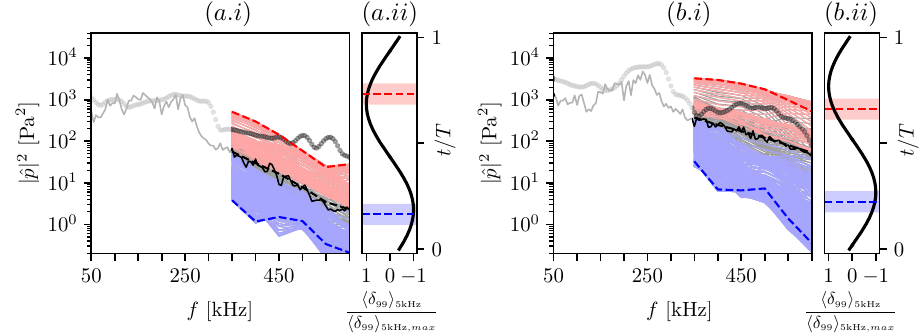}
    \caption{Wall-pressure spectra at sensors ($a$) $s_{6}$ and ($b$) $s_7$.
    (\textit{i}): Symbols ({\protect\colorcircle{1}{1}{1}{0}{0}{0}}) are experimental measurements. Black solid ({\protect\solidLine{0}{0}{0}}) lines are spectra of the assimilated state $\boldsymbol{q} = \mathcal{N}(\boldsymbol{c}_{4})$. Gray lines ({\protect\solidLine{0.65}{0.65}{0.65}}) are $350$ spectra computed using a Hann window of width $T/10=1/(50 \,\mathrm{kHz})$, shifted in steps of $1/(1.75\,\mathrm{MHz})$, and the black dashed line ({\protect\dashedLine{0}{0}{0}})  is their average. 
    Red ({\protect\solidLine{1}{0}{0}}) and blue ({\protect\solidLine{0}{0}{1}}) solid lines are subsets with intensities $\sum_{350\,\mathrm{kHz}}^{600\,\mathrm{kHz}} |\hat{p}|^2$ higher and lower than the average ({\protect\dashedLine{0}{0}{0}}). 
    Red ({\protect\dashedLine{1}{0}{0}}) and blue ({\protect\dashedLine{0}{0}{1}}) dashed lines are the curves with maximum and minimum intensities.
    (\textit{ii}): Black solid ({\protect\solidLine{0}{0}{0}}) lines are normalized $\langle\delta_{\scriptscriptstyle 99}\rangle_{5\,\mathrm{kHz}}$ at the sensors.  Red ({\protect\dashedLine{1}{0}{0}}) and blue ({\protect\dashedLine{0}{0}{1}}) dashed lines are the centers of the Hann windows, $t_0 + T/20$, associated with the identified extrema in (\textit{i}), and the shaded width is $T/10$.}
    \label{fig:fig15_1kHz}
    \vspace*{-4pt}
\end{figure}

We now return to the wall-pressure spectra of the last two sensors, where the high frequencies were under-estimated by the assimilated flow, and we examine the influence of the low-frequency ($5\,\mathrm{kHz}$) flow unsteadiness on these data.  We recompute the wall-pressure spectra during an interval that is one tenth of the slow timescale, $T/10 = 0.02\,\mathrm{ms}=1/(50\,\mathrm{kHz})$. A Hann window is adopted due to the lack of periodicity and to reduce spectral leakage. This short window captures only timescales more than an order of magnitude faster than the $5\,\mathrm{kHz}$ cycle and, as such, the slow variation in the flow can be treated as effectively quasi-steady within the sub-interval. 
A total of $350$ spectra were computed by sliding the Hann window in steps of $1/(1.75\,\mathrm{MHz})$ over one representative $5\,\mathrm{kHz}$ cycle of the wall pressure signal.
The resulting spectra at sensors $s_6$ and $s_7$ are reported in figure~\ref{fig:fig15_1kHz}$(a.i,b.i)$.  While the average of the $350$ spectra (dashed black) recovers the assimilated values (solid black), the ensemble exhibits roughly two orders of magnitude spread, with values that can reproduce or exceed the experimental measurements.  Each curve represents the high-frequency content ($f>300\,\mathrm{kHz}$) at a specific phase in the $5\,\mathrm{kHz}$ cycle, revealing clear intervals where amplification lies entirely above (red) or below (blue) the assimilated value.  The more energetic spectra occur when the boundary layer is larger than its average thickness; the opposite holds for the less energetic spectra. Figures~\ref{fig:fig15_1kHz}$(a.ii,b.ii)$ show that the most energetic pressure spectra (dashed red) are recorded near the peak of the $\langle \delta_{\scriptscriptstyle 99}\rangle_{5\, \mathrm{kHz}}$ oscillation, when the boundary layer is near its thickest. Conversely, the least energetic pressure spectra (dashed blue) occur near the minimum boundary-layer thickness. Notably, these extrema appear at $t \approx 0.75\,T$ and $t\approx 0.25\,T$, respectively, consistent with the spectrally filtered wall-pressure signals in figure~\ref{fig:fig14_1kHz}($c$).

\begin{figure}
    \centering
    \includegraphics[width=0.98\textwidth]{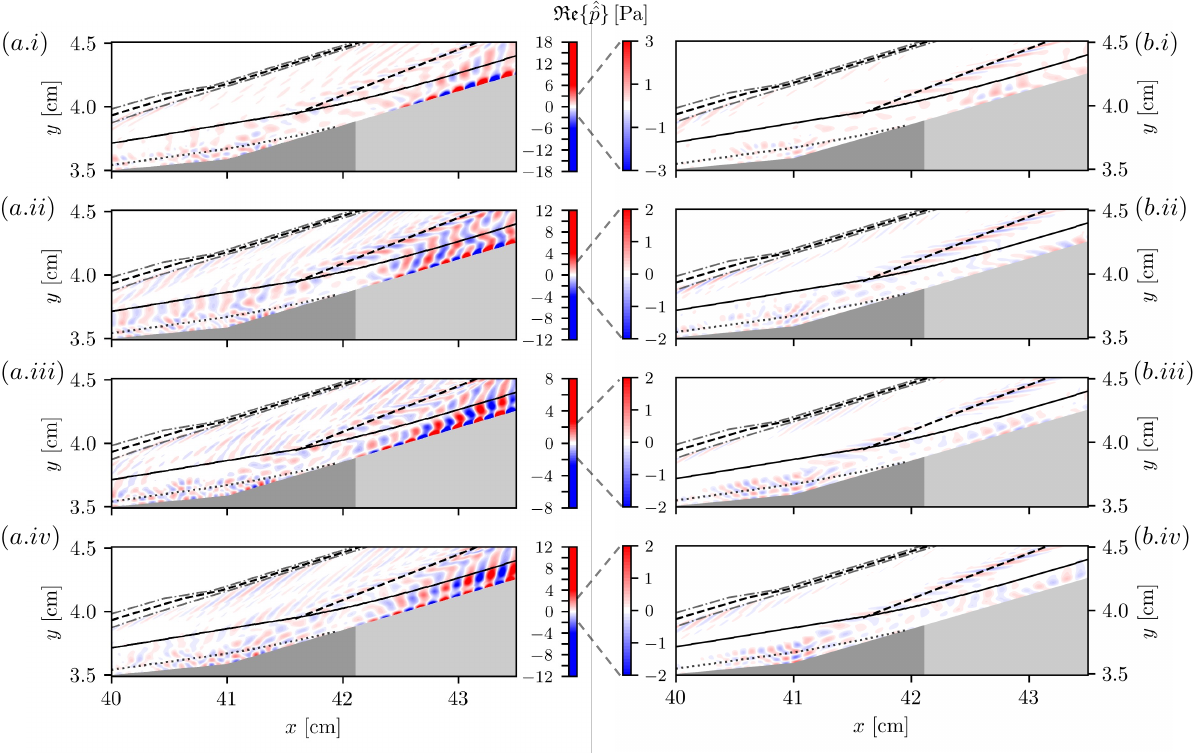} 
    \caption{Pressure Fourier modes from the assimilated state, at frequency $f=450\,\mathrm{kHz}$ and (\textit{i--iv}) $k=\{0, 20, 30, 40\}$. 
    A Hann window is adopted with size $T/10$, over the interval $[t_0, t_0+T/10)$. Choice of $t_0$ in ($a$) maximizes the high-frequency spectra at sensor $s_7$ (red shaded area in figure~\ref{fig:fig15_1kHz}$(b.ii)$); choice of $t_0$ in ($b$) minimizes the high-frequency spectra at sensor $s_7$ (blue shaded area in figure~\ref{fig:fig15_1kHz}$(b.ii)$). (\protect\solidLine{0}{0}{0})  $\overline{\delta}_{\scriptscriptstyle 99}$; (\protect\dashedLine{0}{0}{0}) $\Upsilon(x,y) = 1$; (\protect\dashedLine{0.4}{0.4}{0.4}) $\Upsilon(x,y) = 0.5$; (\protect\dottedLine{0}{0}{0}) $u_{\xi} = 0$; dark and light gray areas as in (\textit{a}); light gray area denotes the separated region, $x=[39.4,\,42.1]~\mathrm{cm}$.}
    \label{fig:fig16_1kHz}
    \vspace*{-4pt}
\end{figure}

Sample Fourier modes, at $f=450\,\mathrm{kHz}$ and $k=\{0, 20, 30, 40\}$, are shown in figure \ref{fig:fig16_1kHz}.  Panels ($a$) and ($b$) contrast the two extreme cases at the last sensor, when the wall pressure fluctuations are large and small (red and blue dashed curves in figure~\ref{fig:fig15_1kHz}$(b.ii)$).  
These results confirm that the behavior is robust across azimuthal wavenumbers, namely that the high-frequency modes near and post reattachment undergo cycles of intensification and weakening, during the low-frequency oscillation of the flow.

In summary, the mismatch between the assimilated flow and the experimental measurements at the final two sensors can be caused by two factors. Firstly, uncertainties in the upstream flow amplify with downstream distance (see error bars in figure \ref{fig:fig08}($a$)), and become largest at sensor $s_7$. In addition, specific to sensor $s_6$, this probe lies within the uncertainty band of the reattachment location, placing it across the transition between separated and reattached flow.
Secondly, the low-frequency unsteadiness of the flow impacts the amplification of the high-frequency boundary-layer disturbances, which have a significant impact on the spectra at the last two downstream sensor locations.

\section{Conclusion}
\label{sec:conclusion}

Transitional, high-speed flow over a cone-flare geometry is simulated.  Unique to the present work is the assimilation of experimental measurements into the direct numerical simulations (DNS), which is achieved using an ensemble-variational (EnVar) approach. In the experiment, Mach 6 flow is established in the reflected-shock tunnel. The test article is a $5^{\circ}$ half-angle cone with a $10^{\circ}$ flare \citep{butler2021_secondmode}.  The experimental measurements consisted of wall-pressure spectra and intensities from seven PCB sensors.  The sensor locations are upstream, within, and downstream of the separation region. The data assimilation attempts to determine the spectra of the incoming boundary-layer instability waves, at the inlet of the simulation domain, that reproduce the experimental measurements.  The fluid dynamical interest is to explain the impact of the flow features, e.g.\,the corner shock and the onset of boundary-layer separation on the assimilated state.

Starting from the measurements at the first sensor, we computed a physics-based initial estimate of the amplitudes of the inflow instability waves, using the linearized Navier-Stokes equations.
When adopted in DNS, the linear estimate accurately reproduces the data at the first sensor, but over-predicts the spectral peak and intensity at the second sensor.  The nonlinear EnVar assimilation procedure is adopted to improve the initial estimate.  Two assimilation tasks are considered:  In the first task, the data from the first two sensors only are used in the assimilation.  The intent is to examine whether closely reproducing these early measurements, where the disturbance dynamics are already nonlinear, is sufficient to accurately predict the downstream state of the flow.  The final estimate of the inflow condition is then adopted as the initial guess in a second assimilation task, where the data from all seven sensors are used.  

The outcome of the first assimilation was instructive. The EnVar procedure altered the spectra of the inflow disturbances relative to the initial guess, by slightly reducing the energy of unstable modes and increasing that of stable ones.   As a result, the wall-pressure intensity was increased at the inflow, did not compromise the accuracy at the first sensor, and eliminated the over-prediction at the second one.  Most importantly, the downstream evolution of this assimilated state, which accurately reproduced the measurements of the first two sensors, deviates appreciably from the downstream sensor data.  The disagreement is not a matter of exponential divergence of trajectories, but rather one of observability: the first two sensors do not observe features of the inflow that are essential to reproduce the downstream dynamics.

The second assimilation that incorporates measurements from all seven sensors further refined the spectra of the inflow disturbances. The adjustments had an appreciable impact on the accuracy of reproducing the wall-pressure intensity at sensors three and four.  These two sensors straddle the onset of separation, which is accurately predicted.  The assimilated flow reproduces the intense rope-like structures characteristic of the upstream attached boundary layer.  When the boundary-layer disturbances reach the separation shock, part of the energy is radiated along the shock as observed in the experiments \citep{butler2021_secondmode}.  Additionally, a significant amplification of boundary-layer disturbances beneath the separation shock takes place.  This effect is undetected in the experimental measurements due to the streamwise  spacing of the PCB probes.  The amplification is primarily observed for the planar waves, which are subsequently quickly attenuated upon entering the separation bubble.  In this region, low-frequency three-dimensional waves amplify, consistent with our and previous linear analyses \citep{paredes2022_cone,dwivedi2019_reattachment}. 

The results show that the measurements downstream of separation are challenging to reproduce, in particular the high frequency components of the wall-pressure spectra at the last two sensors.  Two effects contribute to this difficulty.  
Firstly, uncertainties in the inflow disturbances, when propagated downstream using the Navier-Stokes equations, lead to uncertainties in the high-frequency wall-pressure spectra at these two sensors.  
Secondly, the low-frequency unsteadiness in the separation shock leads to thinning and thickening of the boundary layer, streamwise undulation in the separation and reattachment points, and appreciable changes in the energy of high-frequency disturbances ($f > 300\,\mathrm{kHz}$) at the last two sensors .

Overall, the findings demonstrate the capacity of data assimilation to interpret wall-pressure measurements in high-speed flows that feature shock-boundary layer interaction
and separation.  Since data assimilation is a nonlinear optimization problem, the estimated field depends on the assimilation algorithm, the observability of the available measurements, and the parameterization of the control vector.  Our choice of ensemble-variational assimilation is well suited for statistical measurements. Through the assimilation, we demonstrated the role of upstream sensors on the cone and of probes that straddle separation onset. Future work should examine and contrast the domains of dependence \citep{wang2025DOD} of these sensors and also those near reattachment. Additionally, rigorous approaches to improve the sensor placement and  reduce the uncertainty in the estimated state should be explored. For the parameterization of the control vector, we adopted a superposition of linear instability modes, within the upstream boundary layer.  Future work can evaluate other parameterizations, e.g.\,resolvent modes, in particular far upstream in the early receptivity stages.  Alternatively, data assimilation can be adopted to estimate the flow conditions upstream of the leading-edge shock which can be, for example, expressed in terms of vortical, entropic and acoustic disturbances.


    \par\bigskip
    \noindent
    \textbf{Funding.} The authors acknowledge financial support from the Air Force Office of Scientific Research (grant FA9550-25-1-0011) and the Office of Naval Research (grant N000142512170).  

    \par\medskip
    \noindent
    \textbf{Declaration of interests.} 
    The authors report no conflict of interest.
     

  \newpage
  \bibliography{biblio}
  \bibliographystyle{jfm}

\end{document}